\documentclass[]{aa}

\usepackage{natbib}
\usepackage[leftcaption]{sidecap}
\bibpunct{(}{)}{;}{a}{}{,}
\usepackage{psfig}
\usepackage{rotating}
\usepackage{latexsym}
\def\cm2{${\rm cm^{-2}}$}
\def\mum{${\rm \mu m}$}
\def\ca{$\mathcal{A}$}
\def\cb{$\mathcal{B}$}
\def\cc{$\mathcal{C}$}

\begin{document}

\title{The rich 6 to 9 \mum\, spectrum of interstellar PAHs \thanks{Based on
    observations with ISO, an ESA project with instruments funded by
    ESA Member States (especially the PI countries: France, Germany,
    the Netherlands and the United Kingdom) and with the participation
    of ISAS and NASA.}} 

   \author{E. Peeters\inst{1,2}
          \and S. Hony\inst{3} 
          \and C. Van Kerckhoven\inst{4} 
          \and A.G.G.M. Tielens\inst{2,1} 
          \and L.J. Allamandola\inst{5}
          \and D.M. Hudgins\inst{5}  
          \and C.W. Bauschlicher\inst{5}  }

\offprints{E. Peeters (peeters@astro.rug.nl)}

\institute{SRON National Institute for Space Research, P.O. Box 800,
  9700 AV Groningen, The Netherlands
   \and Kapteyn Institute, P.O. Box
  800, 9700 AV Groningen, The Netherlands 
  \and Astronomical Institute
  ``Anton Pannekoek'', Kruislaan 403, 1098 SJ Amsterdam, The
  Netherlands 
  \and Instituut voor Sterrenkunde, K.U.Leuven,
  Celestijnenlaan 200B, 3100 Heverlee, Belgium 
  \and NASA-Ames Research
  Center, Space Science Division, MS: 245-6, Moffett Field, CA
  94035-1000, U.S.A.}

\date{}
\titlerunning{The rich 6 to 9 \mum\, spectrum of interstellar PAHs}
 
\authorrunning{E. Peeters}

\abstract{IR spectroscopy provides a valuable tool for the
  characterisation and identification of interstellar molecular
  species. Here, we present 6--9 \mum\, spectra of a sample of
  reflection nebulae, HII regions, YSOs, evolved stars and galaxies
  that show strong unidentified infrared bands, obtained with the SWS
  spectrograph on board ISO. The IR emission features in this
  wavelength region show pronounced variations. 1) The 6.2 \mum\,
  feature shifts from 6.22 to 6.3 \mum\, and clearly shows profile
  variations. 2) The 7.7 \mum\, complex is comprised of at least two
  subpeaks peaking at 7.6 and one longwards of 7.7 \mum. In some cases
  the main peak can apparently shift up to 8 \mum. Two sources do not
  exhibit a 7.7 \mum\, complex but instead show a broad emission
  feature at 8.22 \mum. 3) The 8.6 \mum\, feature has a symmetric
  profile in all sources and some sources exhibit this band at
  slightly longer wavelengths.  For the 6.2, 7.7 and 8.6 \mum\,
  features, the sources have been classified independently based on
  their profile and peak position. The classes derived for these
  features are directly linked with each other. Sources with a 6.2
  \mum\, feature peaking at $\sim$ 6.22 \mum\, exhibit a 7.7 \mum\,
  complex dominated by the 7.6 \mum\, component. In contrast, sources
  with a 6.2 \mum\, profile peaking longwards of 6.24 \mum\, show a
  7.7 \mum\, complex with a dominant peak longwards of 7.7 \mum\, and
  a 8.6 \mum\, feature shifted toward the red. Furthermore, the
  observed 6--9 \mum\, spectrum depends on the type of object. All
  ISM-like sources and a few PNe and Post-AGB stars belong to the
  first group while isolated Herbig AeBe stars, a few Post-AGB stars
  and most PNe belong to the second group.  We summarise existing
  laboratory data and theoretical quantum chemical calculations of the
  modes emitting in this wavelength region of PAH molecules. We
  discuss the variations in peak position and profile in view of the
  exact nature of the carrier. We attribute the observed 6.2 \mum\,
  profile and peak position to the combined effect of a PAH family and
  anharmonicity with pure PAHs representing the 6.3 \mum\, component
  and substituted/complexed PAHs representing the 6.2 \mum\,
  component. The 7.6 \mum\, component is well reproduced by both pure
  and substituted/complexed PAHs but the 7.8 \mum\, component remains
  an enigma. In addition, the exact identification of the 8.22 \mum\,
  feature remains unknown. The observed variations in the
  characteristics of the IR emission bands are linked to the local
  physical conditions. Possible formation and evolution processes that
  may influence the interstellar PAH class are highlighted.
  \keywords{Circumstellar matter - Stars: pre-main sequence - HII
    regions - ISM: molecules; - Planetary nebulae: general - Infrared:
    ISM: lines and bands} } \maketitle

\section{Introduction}

Mid-infrared spectra of many sources are dominated by the well-known
emission features at 3.3, 6.2, 7.7 and 11.2 \mum, commonly called the
unidentified infrared (UIR) bands \citep[cf.][]{Gillett:73,
  Geballe:85, Cohen:co:86}. These UIR bands are associated with a wide
variety of objects - including HII regions, Post-AGB stars, PNe, YSOs,
the diffuse ISM and galaxies - and are generally attributed
to polycyclic aromatic hydrocarbon (PAH) molecules \citep{Leger:84,
  Allamandola:autoexhaust:85, Puget:revpah:89, Allamandola:rev:89},
although the exact molecular identification of the carriers remains
unknown. Beyond serving as simple PAH indicators, they can serve as
red-shift indicators, as tracers of elemental evolution in external
galaxies, as tracers of chemical evolution and can be used to probe
environmental conditions within the objects \citep{Genzel:ULIRGs:98,
Lutz:ULIRGs:98, Helou:parijs:99, Serabyn:ngst:99, Genzel:rev:00,
Helou:normalgal:00, Joblin:isobeyondthepeaks:00, Hony:oops:01, Vermeij:pahs:01,
Verstraete:prof:01}. 

The region from 6 to 9 \mum\, reveals a number of emission features
with bands at 5.2, 5.7, 6.0, 6.2, 6.8, "7.7" and 8.6 \mum. The "7.7"
\mum\, feature is particularly important as it is the strongest of the 
interstellar "UIR" bands and, as such, can be used to probe
objects in which the other features are weak. 

Until quite recently, most of the interstellar emission bands were
considered to be more-or-less invariant in position and profile. Although
some minor variations were noted, by and large the 6.2 \mum \, feature
was considered fixed at 6.2 \mum, regardless of the reported shift in
peak position by \citet{Molster:pah:96}. The "7.7" \mum \, band was
generally treated similarly in spite of earlier papers showing this
band is comprised of at least two variable components
\citep[e.g.][]{Bregman:hiivspn:89, Cohen:southerniras:89,
  Beintema:pahs:96, Molster:pah:96, Roelfsema:pahs:96, Moutou:leshouches:99,
  Moutou:parijs:99, Peeters:parijs:99}.
It was recognised some time ago that the 7.7 \mum\, complex appears
either with a dominant 7.6 \mum\, component or with the dominant
component peaking at 7.8--8 \mum\,
\citep{Bregman:hiivspn:89,Cohen:southerniras:89}. In 
addition, it was found that the former profile is associated with
HII regions and the one peaking near 7.8 \mum \, is associated with
planetary nebulae
\citep{Bregman:hiivspn:89,Cohen:southerniras:89}. Recently, thanks to 
the high resolution spectra obtained with ISO, more
subpeaks of the 7.7 \mum\, complex were reported near 7.2 to 7.4
and 8.2 \mum\, \citep{Moutou:leshouches:99, Moutou:c60:99}. 

In Sect. \ref{data}, our sample and the observations are presented,
the data reduction, the influence of extinction and the decomposition
of the spectra are discussed. Sections \ref{f62}, \ref{f77} and
\ref{f86} analyse the 6.2, 7.7 and 8.6 \mum\, features respectively.
The link between the observed variations in the 6.2, 7.7 and 8.6
\mum\, features and the connection with the type of object is
highlighted in Sect. \ref{classes}. Section \ref{corr} presents the
observed trends. The spectral characteristics of PAHs in this
wavelength range as measured in the laboratory and calculated by
quantum chemical theories are summarised in Sect. \ref{spec}. Section
\ref{discussion} highlights the astronomical implications. Finally, in
Sect. \ref{summary} our main results are summarised.

\section{The data}
\label{data}

\subsection{Sample}

\begin{table*}[!th]
\caption[The sample]{Journal of observations. The coordinates of the
  SWS pointing are given.} 
\label{logbook}
  \begin{center}
    \leavevmode
    \scriptsize
{\setlength{\tabcolsep}{0.16cm}
\begin{tabular}{llrlllclll}\\[-0.3cm]
\hline \\[-0.1cm]
Source           & \multicolumn{1}{c}{$\alpha$} & \multicolumn{1}{c}{$\delta$} &
\multicolumn{1}{c}{TDT$^b$} & Obs. &  Ref. & A$_K ^d$ & Sp. Type$^d$ &
G$_0 ^d$  & \multicolumn{1}{c}{Object Type}    \\
                 & \multicolumn{1}{c}{(J2000)$^a$} &
\multicolumn{1}{c}{(J2000)$^a$}&
\multicolumn{1}{c}{ } & mode$^c$ &   & & & &     \\[0.1cm] 
\hline\\[-0.1cm]         
NGC~253           & 00 47 33.19 & $-$25 17 17.20 & 24701422 & 01(4) & 1     &   &  -     & -   & Seyfert Galaxy \\[0.03cm]
W~3A 02219+6125   & 02 25 44.59 & $+$62 06 11.20 & 64600609 & 01(2) & 2     & 1.5 &  O6    & 1E4 & CHII\\[0.03cm]
IRAS~02575+6017   & 03 01 31.28 & $+$60 29 13.49 & 15200555 & 01(2) & 2     & 2   &        & 1E5 & CHII+YSO\\[0.03cm]
IRAS~03260+3111   & 03 29 10.37 & $+$31 21 58.28 & 65902719 & 01(3) & 3     &   &  B9    & 2E4 & non-isolated Herbig Ae Be stars \\[0.03cm]
Orion PK1         & 05 35 13.67 & $-$05 22 08.51 & 68701515 & 01(4) & 4     & 0.15&  O6    &     & HII \\[0.03cm]
Orion PK2         & 05 35 15.79 & $-$05 24 40.69 & 83301701 & 01(4) & -     &   &  O6    &     & HII \\[0.03cm]
OrionBar~D8       & 05 35 18.22 & $-$05 24 39.89 & 69501409 & 01(2) & 5     &   &  O6    &     & HII \\[0.03cm]
OrionBar~BRGA     & 05 35 19.31 & $-$05 24 59.90 & 69502108 & 01(2) & -     &   &  O6    &     & HII \\[0.03cm]
OrionBar~D5       & 05 35 19.81 & $-$05 25 09.98 & 83101507 & 01(2) & -     &   &  O6    & 5E4 & HII \\[0.03cm]
OrionBar~H2S1     & 05 35 20.31 & $-$05 25 19.99 & 69501806 & 01(4) & 6     &   &  O6    & 7E3 & HII \\[0.03cm]
OrionBar~D2       & 05 35 21.40 & $-$05 25 40.12 & 69502005 & 01(2) & -     &   &  O6    &     & HII \\[0.03cm]
NGC~2023          & 05 41 38.29 & $-$02 16 32.59 & 65602309 & 01(3) & 7     &   &B1.5V   & 3E2 & RN \\[0.03cm]
HD~44179          & 06 19 58.20 & $-$10 38 15.22 & 70201801 & 01(4) & 8     &   & B8V    & 5E6 & Post-AGB star \\[0.03cm]
IRAS~07027-7934   & 06 59 26.29 & $-$79 38 48.01 & 73501035 & 01(2) & 9     &   &WC10    & 2E7 & PN\\[0.03cm]  
M~82              & 09 55 50.70 & $+$69 40 44.40 & 11600319 & 01(4) & 1     &   &  -     & -   & starburst galaxy\\[0.03cm]
HR~4049           & 10 16 07.56 & $-$28 59 31.31 & 17100101 & 01(2) & 8,10  &   &B9.5Ib-II&    & Post-AGB star \\[0.03cm]
IRAS~10589-6034   & 11 00 59.78 & $-$60 50 27.10 & 26800760 & 01(2) & 2     & 1.5 &        & 1E5 & CHII  \\[0.03cm]
HD~97048          & 11 08 04.61 & $-$77 39 18.88 & 61801318 & 01(4) & 11    & 0.12&  A0    & 1.7E4 & non-isolated Herbig Ae Be star \\[0.03cm]
HD~100546         & 11 33 25.51 & $-$70 11 41.78 & 27601036 & 01(1) & 12    & 0.03&  B9Vne & 9E3 & isolated Herbig Ae Be star \\[0.03cm]
IRAS~12063-6259   & 12 09 01.15 & $-$63 15 54.68 & 25901414 & 01(2) & 2     & 1.5 &        & 1E5 & CHII  \\[0.03cm]
IRAS~12073-6233   & 12 10 00.32 & $-$62 49 56.50 & 25901572 & 01(2) & 2     & 1.5 &O6-O7.5 & 1E6 & CHII/star forming region  \\[0.03cm]
IRAS~13416-6243   & 13 46 07.61 & $-$62 58 19.98 & 62803904 & 01(3) & -     &   &        &     & Post-AGB star \\[0.03cm]     
circinus          & 14 13 09.70 & $-$65 20 21.52 & 07902231 & 01(4) & 13    &   &  -     & -   & Seyfert 2 galaxy\\[0.03cm]
HE~2-113          & 14 59 53.49 & $-$54 18 07.70 & 43400768 & 01(2) & 14    &   &  WC10  & 6E4 & PN\\[0.03cm]      
IRAS~15384-5348   & 15 42 17.16 & $-$53 58 31.51 & 29900661 & 01(2) & 2     & 1.5 &        & 5E4 & CHII  \\[0.03cm]
G~327.3-0.5       & 15 53 05.89 & $-$54 35 21.08 & 11702216 & 01(1) & -     &   &38,000  &     & HII \\[0.03cm]
IRAS~15502-5302   & 15 54 05.99 & $-$53 11 36.38 & 27301117 & 01(2) & 2     & 3.1 &        & 3E6 & CHII  \\[0.03cm]
IRAS~16279-4757   & 16 31 38.20 & $-$48 04 06.38 & 64402513 & 01(3) & 15    &   &        &     & Post-AGB star \\[0.03cm]
CD~-42 11721 (off)& 16 59 05.82 & $-$42 42 14.80 & 28900461 & 01(2) &3,16,17& 0.7 &  B0    &     & non-isolated Herbig Ae Be star\\[0.03cm]
CD~-42 11721      & 16 59 06.79 & $-$42 42 07.99 & 64701904 & 01(2) & 16,17 & 0.4-0.7 & B0 &     & non-isolated Herbig Ae Be star\\[0.03cm]
IRAS~17047-5650   & 17 00 00.91 & $-$56 54 47.20 & 13602083 & 01(3) & 9     &   &  WC10  & 5E6 & PN\\[0.03cm]    
IRAS~16594-4656   & 17 03 09.67 & $-$47 00 47.90 & 45800441 & 01(1) & 18    &   &  B7    &     & Post-AGB star \\[0.03cm]
IRAS~17279-3350   & 17 31 17.96 & $-$33 52 49.30 & 32200877 & 01(2) & 2     & 2.2  &        &  5E3& CHII\\[0.03cm]
IRAS~17347-3139   & 17 36 00.61 & $-$31 40 54.19 & 87000939 & 01(3) & 19    &   &        & 8E5 & PN\\[0.03cm]       
XX-OPH            & 17 43 56.42 & $-$06 16 08.00 & 46000601 & 01(4) & -     &   &  Ape   &     & variable star, irregular type\\[0.03cm] 
Hb 5              & 17 47 56.11 & $-$29 59 39.70 & 49400104 & 01(3) & 17    &   &120,000 &     & PN\\[0.03cm]      
IRAS~18032-2032   & 18 06 13.93 & $-$20 31 43.28 & 51500478 & 01(2) & 2     & 1.1  &        & 2E5 & CHII  \\[0.03cm]
IRAS~18116-1646   & 18 14 35.29 & $-$16 45 20.99 & 70300302 & 06    & 2     &   &        & 8E4 & CHII\\[0.03cm]
GGD~-27 ILL       & 18 19 12.03 & $-$20 47 30.59 & 14900323 & 01(2) &       & 2   &  B1    & 1E6 & star forming region \\[0.03cm]
                  & 18 19 12.00 & $-$20 47 31.10 & 14802136 & 01(2) & 2,20  &     &        &   &  \\[0.03cm]
MWC~922           & 18 21 16.00 & $-$13 01 30.00 & 70301807 & 01(2) & -     &  &  Be    & 6E6 & emission-line star\\[0.03cm]
IRAS~18317-0757   & 18 34 24.94 & $-$07 54 47.92 & 47801040 & 01(2) & 2     & 2.0 &  O8    & 1E5 & CHII  \\[0.03cm] 
IRAS~18434-0242   & 18 46 04.09 & $-$02 39 20.02 & 51300704 & 06    & 21    & 1.6 &  O3-O5 & 2E6 & CHII\\[0.03cm]
IRAS~18502+0051   & 18 52 50.21 & $+$00 55 27.59 & 15201645 & 01(2) & 2     &   &  O7    & 1E6 & CHII  \\[0.03cm] 
HD~179218         & 19 11 11.16 & $+$15 47 18.58 & 32301321 & 01(3) & 22    & 0.37&  B9    &$>$2E4& isolated Herbig Ae Be star \\[0.03cm]
IRAS~18576+0341   & 19 00 10.50 & $+$03 45 47.99 & 32401203 & 01(1) & 23    &   &15,000  & 4E3 & LBV \\[0.03cm]  
BD~+30 3639       & 19 34 45.19 & $+$30 30 58.79 & 86500540 & 01(3) & 14    &   &  WC9   & 1E5 & PN\\[0.03cm]      
IRAS~19442+2427   & 19 46 20.09 & $+$24 35 29.40 & 15000444 & 01(2) & 2,20  &   &  O7    & 7E6 & CHII  \\[0.03cm]
BD~+40 4124       & 20 20 28.31 & $+$41 21 51.41 & 35500693 & 01(3) & 24    & 0.3 &  B2V   & 1E4 & non-isolated Herbig Ae Be star  \\[0.03cm]
S~106 (IRS4)      & 20 27 26.68 & $+$37 22 47.89 & 33504295 & 01(2) & 25    & 1.4 &  O8    & 2E5 & YSO \\[0.03cm]  
NGC~7023 I        & 21 01 31.90 & $+$68 10 22.12 & 20700801 & 01(4) & 7     &   &  B3    & 5E2 & RN \\[0.03cm]
CRL~2688          & 21 02 18.79 & $+$36 41 37.79 & 35102563 & 01(3) & -     &   &  F5Iae & 5E3 & Post-AGB star \\[0.03cm]      
NGC~7027          & 21 07 01.70 & $+$42 14 09.10 & 55800537 & 01(4) & 8     &   & 200,000& 2E5 & PN\\[0.03cm]      
IRAS~21190+5140   & 21 20 44.89 & $+$51 53 26.99 & 74501203 & 06    & 21    & 0.0 &        & 6E5 & CHII\\[0.03cm]
IRAS~21282+5050   & 21 29 58.42 & $+$51 03 59.80 & 05602477 & 01(2) & 8     &   &  O9    & 1E5 & Post-AGB star \\[0.03cm]
IRAS~22308+5812   & 22 32 45.95 & $+$58 28 21.00 & 17701258 & 01(2) & 2,20  &   &  O7.5  & 1E3 & CHII  \\[0.03cm] 
IRAS~23030+5958   & 23 05 10.60 & $+$60 14 40.99 & 75101204 & 06    & -     & 0.0 &  O6.5  & 8E3 & CHII\\[0.03cm]
IRAS~23133+6050   & 23 15 31.39 & $+$61 07 08.00 & 56801906 & 01(2) & 2     & 0.4 &  O9.5  & 3E5 & CHII  \\[0.2cm] 
\hline\\[-0.5cm]
\end{tabular} }
\end{center}
$^{a}$ : Units of $\alpha$ are hours, minutes, and seconds, and units
of $\delta$ are degrees, arc minutes, and arc seconds. 
\,  $^{b}$ : each ISO observation is given a unique TDT (Target
Dedicated Time) number. \, $^c$ SWS observing mode used
\citep[see][]{deGraauw:sws:96}. Numbers in brackets correspond to the scanning
speed. \, $^d$ see text for details.\\
References : 1 : \citet{Sturm:swsgal:00}; 2 : \citet{Peeters:cataloog:01}; 3 : \citet{VanKerckhoven:plat:00}; 
4 : \citet{Rosenthal:parijs:99}; 5 : \citet{Cesarsky:sileminorion:00}; 6 : \citet{Verstraete:prof:01}; 
7 : \citet{Moutou:leshouches:99}; 8 : \citet{Beintema:pahs:96}; 9 : \citet{Szczerba:pahinwrpn:01}; 
10 : \citet{Molster:pah:96}; 11 : \citet{VanKerckhoven:parijs:99}; 12 : \citet{Waelkens:hd10:96}; 
13 : \citet{Moorwood:circinus:96}; 14 : \citet{Waters:crystsilpn:98}; 15 : \citet{Tielens:parijs:99}; 
16 : \citet{Benedettini:98}; 17 : \citet{Hony:oops:01}; 18 : \citet{Garcia:16594:99}; 
19 : \citet{Cohen:dustin17347:99}; 20 : \citet{Roelfsema:pahs:96} ; 21 : \citet{Peeters:parijs:99}; 
22 : \citet{Meeus:herbigaebe:01}; 23 : \citet{Hrivnak:swsPPN:00}; 24 : \citet{Wesselius:bd40:96}; 
25 : \citet{vandenAncker:s106andcep:00} 
\end{table*}

The sample includes 57 sources from a wide variety of objects, ranging
from Reflection Nebulae (RNe), HII regions, Young Stellar Objects
(YSOs), Post-AGB stars, Planetary Nebulae (PNe) to galaxies (see
Table \ref{logbook}).  We give in Table \ref{logbook}
characteristics of the sources; i.e. the extinction in the K-band, A$_K$, the
spectral type of the illuminating source, and an estimate of the
incident UV flux density at 1000\AA, $G_0$, at the location where the PAH
emission originates in units of the average interstellar radiation field
\citep[$\lambda u_{\lambda}$=1.6 10$^{-6}$\, W/m$^2$,][]{Habing:G0:68}.

For the compact HII regions (CHII) present in our sample, \citet{paperii}
estimated A$_K$ based upon HI recombination lines to be between 0 and
2.7 magn. A$_K$ is taken from \citet{Cidale:bestars_parameters:01} for
CD~-42~11721, from \citet{Miroshnichenko:haebe:99} for HD~179218, from
\citet{Everett:peak1:95} for Orion Peak1 and from
\citet{vandenAncker:thesis:99} for IRAS~03260, GGD~-27, S~106,
HD~97048, BD~+40~4124 and HD~100546.

For most sources, the spectral types are taken from Simbad. The
spectral type of IRAS~12073 and IRAS~18434 are taken from \citet[][and
private communication]{Kaper:boulder:02,Kaper:vlt:02} and that of
IRAS~16594-4656 is from \citet{Su:imagingPPN:01}. The effective
temperatures for Hb 5, NGC7027, IRAS~18576 and G327 are taken from
\citet{Gesicki:Teff:00}, \citet{Latter:Teffngc7027:00},
\citet{Ueta:18576:01} and \citet{Ehrenfreund:swsmassstars:97}
respectively.

For the CHII regions and GGD~-27 ILL, we have derived $G_0$ values
from the observed IR flux and the angular size of the PAH emission
region \citep[cf. ][]{Hony:oops:01}. This estimate is based on the
assumption that all the UV light is absorbed in a spherical shell with
the angular diameter of the HII region and re-emitted in the IR. We
have used for the size of the HII regions the measured radio sizes.
This is reasonable since the PAHs are expected to be destroyed inside
the HII region.  The IR flux was derived from the L$_{IR}$ given by
\citet{Peeters:cataloog:01} and the radio sizes used are taken from
\citet{Peeters:cataloog:01} and \citet{Martin:radio:02}. The G$_0$
values are similar to those derived by \citet{Hony:oops:01} for the
sources present in both samples. For the Orion bar, we refer to
\cite{Tielens:anatomyorionbar:93} and \citet{Joblin:3umvsmethyl:96}
for the given $G_0$ values. We have taken G$_0$ values for the Herbig
Ae Be stars from Van Kerckhoven et al. (2002, in preparation) who
derived $G_0$ from the UV flux between 6 and 13.6 eV, F$_{UV}$, and
the spatial distribution of the PAHs in the sources. F$_{UV}$ is
derived from the observed stellar flux and the known spectral type.
CRL~2688 has an effective temperature of $\sim$6400K. Hence, the FUV
luminosity is 0.04\% of the total luminosity of the star. The star's
luminosity and the NIR size are taken from \citet{Goto:crl2688:02}.
The FIR flux of IRAS~17347 and IRAS~18576 are obtained by integrating
the modified blackbody that is fitted to the SWS spectra. The size of
IRAS~17347 and IRAS~18576 are taken from \citet{Meixner:imppn:99} and
\citet{Ueta:18576:01} respectively. For MWC~922, the diameter is taken
from \citet{Meixner:imppn:99} and its FIR flux is derived by
integrating the combined SWS and LWS spectrum longwards of 20 \mum.
For the RNe, PNe and Post-AGB stars not mentioned in this paragraph,
the $G_0$ values are taken from \citet{Hony:oops:01}.

\subsection{Observations}

All spectra presented here were obtained with the Short Wavelength
Spectrometer \citep[SWS,][]{deGraauw:sws:96} 
on board the Infrared Space Observatory \citep[ISO,][]{Kessler:iso:96}. The spectra were 
taken using the AOT~01 scanning mode at various speeds or the AOT~06 mode,
with resolving power ($\lambda/\Delta \lambda$) ranging from
500 to 1600. See Table \ref{logbook} for details of the observations.

\subsection{Reduction}

The data were processed with the SWS Interactive Analysis package
IA$^{3}$ \citep{deGraauw:sws:96} using calibration files and
procedures equivalent with pipeline version 7.0 or later. Further data
processing consisted of bad data removal and rebinning with a constant
resolution. The (sub-)features discussed here are present in all
available scans.

\begin{figure}[!hbt]
  \begin{center}
   \begin{picture}(240,240)(-5,10)
   \psfig{figure=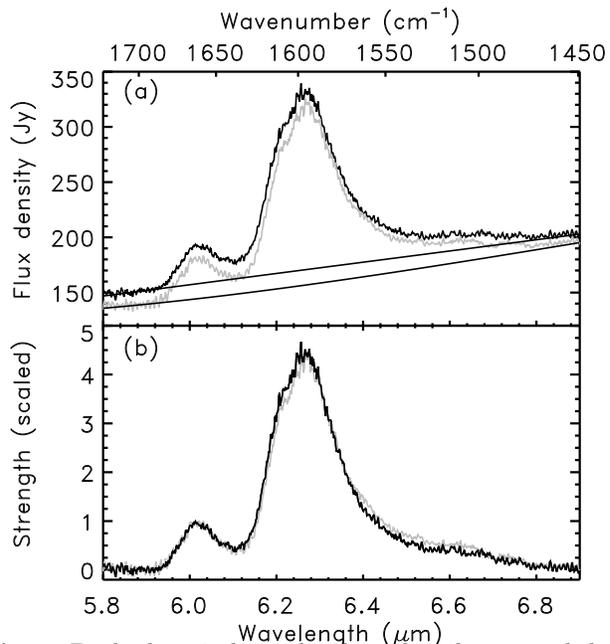,width=240pt,height=240pt}
   \end{picture}
  \end{center}
\caption{Both the - independently reduced - up and down scans of
  HD~44179 are shown with their respective continua in panel a. Panel b
  shows the normalised profiles of the up and down scan.}
\label{invl_memeff62}
\end{figure}

\begin{figure}[!bht]
  \begin{center}
   \begin{picture}(240,240)(-5,10)
   \psfig{figure=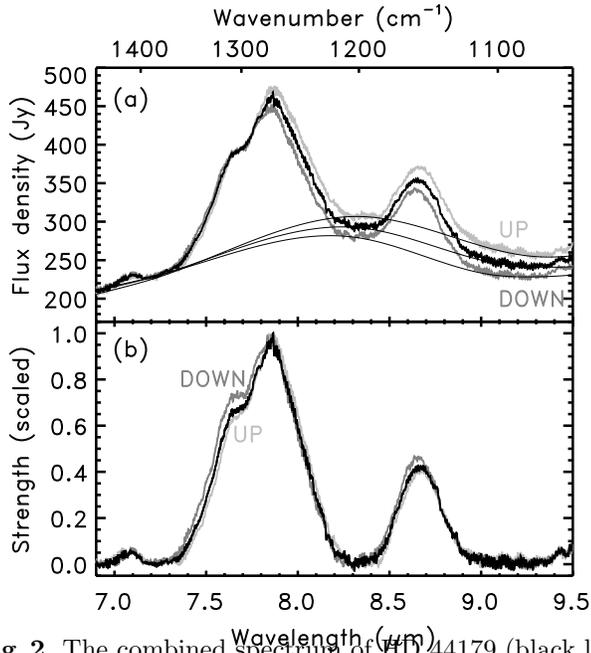,width=240pt,height=240pt}
   \end{picture}
  \end{center}
\caption{The combined spectrum of HD~44179 (black line) together with the -
  independently reduced - up and down scans are shown with their
  respective continua in panel a. The continuum subtracted profiles
  are shown in panel b normalised to the peak intensity.}
\label{invl_memeff77}
\end{figure}

In case of high fluxes, the obtained spectra can suffer from memory
effects. These memory effects can influence the general shape of the
continuum as well as the profile of broad features. The sources in our
sample for which memory effects are present, are indicated in Table
\ref{sample}. At the time the data reduction was done, no memory
correction tool was available. Hence, in case of memory effects, the
average of the up and down scans is taken. In order to investigate
the influence of memory effects on this study, we analyse the source
that suffers the most from memory effects in our sample, i.e.
HD~44179, by comparing the up and down scans in the region of interest,
i.e. 5.5-9 \mum.  Fig. \ref{invl_memeff62} shows the influence on the
6.2 \mum\, profile. The differences are small, even in this most
extreme case. The influence is more severe for the 7.7 \mum\,complex
(Fig. \ref{invl_memeff77}). The blue wing of the feature is affected,
as well as the relative strength of the 7.6 component. However, the
error due to detector memory effects ($<$  5 \%) is less
  than the uncertainty on the integrated band intensity. It
  will not hamper the spectral analysis and source classification
  performed in this paper. Hence, it will
  not hamper the analysis done in this paper. Recently, a memory
correction tool has become available (OLP10) and, as a check, the
sources suffering from memory effects have been re-reduced. We found
that that memory effects do not alter significantly the band
profiles. In order to be consistent with the analysis of the other
sources, we did not apply this memory correction.
 
Two sources in this sample (Orion peak 1 and Orion peak 2) have strong
atomic emission lines perched on top of the 6.2 \mum\, PAH
feature. These lines and the PAH feature are easily separated at the
resolution of the SWS instrument. The contribution from any line is
removed prior to the analysis of the profiles.

\subsection{The spectra}
\label{spectra}

\begin{figure}[!t]
  \begin{center}
   \begin{picture}(240,240)(-5,10)
   \psfig{figure=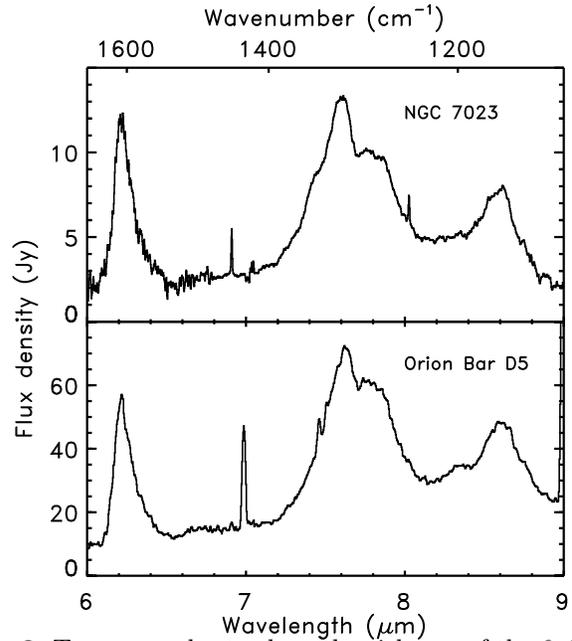,width=240pt,height=240pt}
   \end{picture}
  \end{center}
\caption{Two examples to show the richness of the 6--9 \mum \, region.
  We recognise the 6.2 and 8.6 \mum\, features. The "7.7" \mum\,
  feature breaks up in three components; two clear bands at 7.6 and
  7.8 \mum\, and a shoulder at 7.4 \mum. Furthermore, weak features
  are present at 6.6 and 8.2 \mum.}
\label{prof6_fig_spectra}
\end{figure}

Fig. \ref{prof6_fig_spectra} shows spectra of two typical sources to
illustrate the spectral detail present. The complete 6--9 \mum\,
spectrum reveals an extremely rich collection of emission features with
bands at 6.0, 6.2, 6.6, 7.0, "7.7", 8.3 and 8.6 \mum \, \citep{
  Beintema:pahs:96, Molster:pah:96, Roelfsema:pahs:96,
  Verstraete:m17:96, Moutou:leshouches:99,
  Moutou:parijs:99, Peeters:parijs:99, Verstraete:prof:01}. In
particular, upon close inspection, some of
these features are perched on top of an emission plateau of variable
strength. The beginning of this emission plateau seems to be
variable and falls longwards of 6 \mum\, while it extends until $\sim$ 9 \mum.

From the richness of the region, it is clear that several components
are present. The well known 7.7 \mum\, feature consists of two main
features at 7.6 and 7.8 \mum \, plus shoulders at 7.3-7.4, 7.45 and 8.2
\mum. For example, \citet{Verstraete:prof:01} fit the
total region with several Lorentzian profiles and Van Kerckhoven
et~al. (2002, in prep.) fit the 7.7 \mum\, complex with 4 Gaussians
peaking at 7.5, 7.6, 7.8 and 8.0 \mum. 

\subsection{Decomposition of the spectra}

\subsubsection{The continuum}
\label{continuum}

The profile of the 6.2 \mum\, feature in the spectra of all sources is
determined by subtracting a local spline continuum or a polynomial of
order 1.  To assess the sensitivity of the resulting profiles to the
continuum choice, two extreme baselines have been defined and
subtracted.  In general, the influence of the continuum determination
on the profile is very small and hence does not change significantly
the band profiles nor the source classification performed hereafter.
In some sources however, the continuum determination is subject to
some freedom.  These sources are indicated in Table \ref{sample}.
 
\begin{figure}[!htb]
  \begin{center}
   \begin{picture}(240,300)(-5,10)
   \psfig{figure=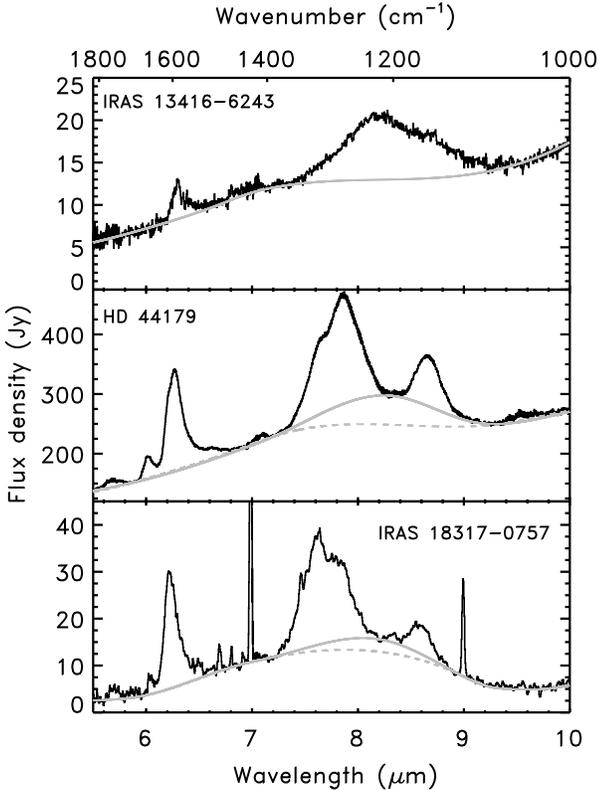,width=240pt,height=300pt}
   \end{picture}
  \end{center}
\caption{Illustrative examples of the continua underneath
  the 7.7 and 8.6 \mum\,features. The dashed line represents the
  general continuum, the full line the second (local) continuum. See
  text for details.}
\label{cont}
\end{figure}

The continuum determination around the 7.7 and 8.6 \mum\,features is
quite arbitrary. We choose to draw first a general continuum splined
through points from 5-6 and 9-10 \mum\, and through points near 7
\mum, excluding possible small features in those regions (see
Fig. \ref{cont}, dashed line). In this way,
the influence of a silicate absorption feature in some sources (see
Table \ref{sample}) is completely ignored.  In addition, to separate
and study the individual 7.7 and 8.6 \mum\, contributions, we have
also drawn a continuum under the 7.7 and 8.6 \mum\, features
themselves. This second (local) continuum is determined by taking
additional continuum points near 8.3 \mum\, - between the 7.7 and 8.6
\mum\, features (see Fig. \ref{cont}, full line). In this way, an underlying
plateau component is defined.

Other ways of decomposing the broad, blended bands and determining the
underlying continuum will yield other results. In particular, for
different bandshapes (Gaussian, Lorentzian, etc.), 
different continua and profile parameters (central wavelength and FWHM)
are obtained \citep{Boulanger:lorentz:98, Uchida:RN:00}. However, these
differences will affect all sources in a systematic way and while this
will influence the profiles of the derived features, this will not
affect the source-to-source variations we follow.

\begin{figure*}[!hbt]
  \begin{center}
   \begin{picture}(540,300)(10,10)
   \psfig{figure=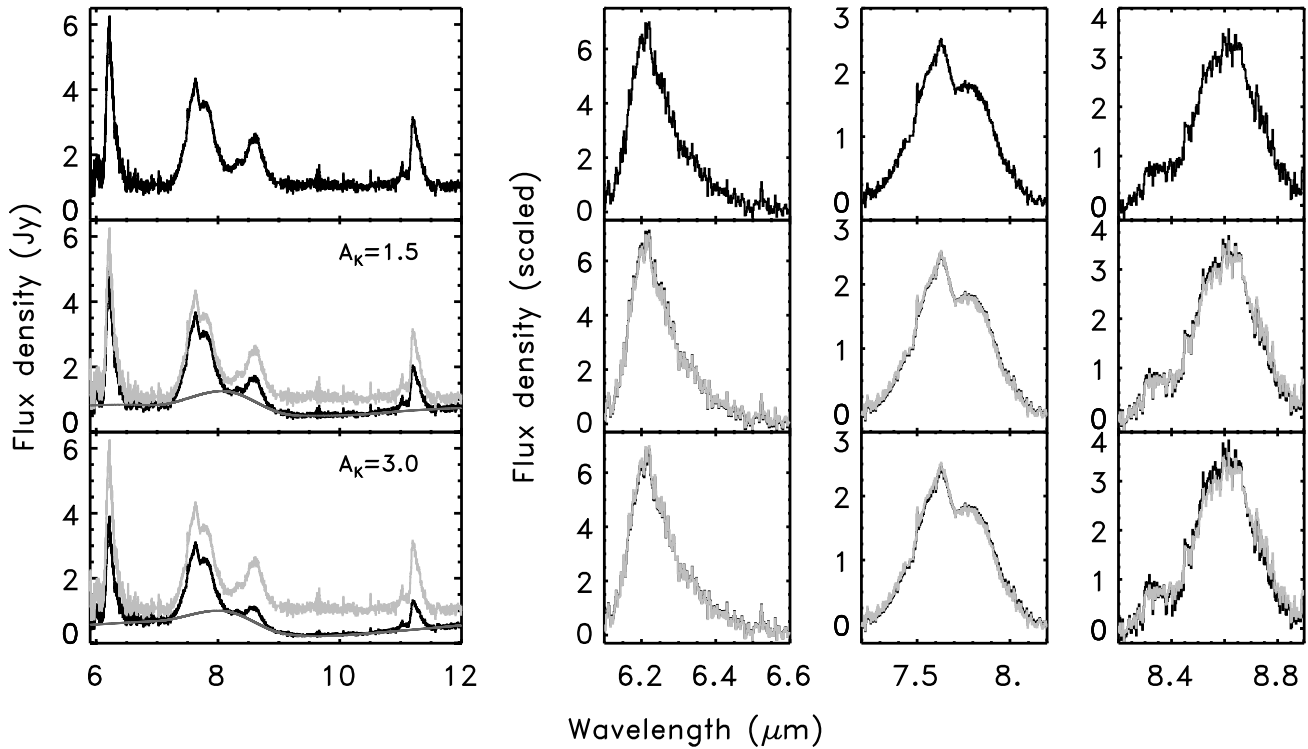,width=540pt,height=300pt}
   \end{picture}
  \end{center}
\caption{ The influence of extinction on a template PAH spectrum. The
    top left panel shows the template, a PAH spectrum on top of a
    continuum of 1. The "standard" extinction law is applied to this
    template spectrum for an A$_K$ of 1.5 and 3.0 (middle and lower
    left panels resp.). As a reference, the template spectrum is
    plotted in grey. In addition, the derived continuum is shown for
    the extincted spectra. The derived profiles are shown in the right
    panels. As a reference, the profiles of the template spectrum are
    plotted in grey on top of the derived profiles of the extincted
    spectra.}
\label{Prof6_ext}
\end{figure*}

\subsubsection{Extinction}
\label{extinction}

Extinction can have a serious effect on the apparent PAH spectrum
\citep[][see also Fig. \ref{Prof6_ext}]{Spoon:silenpahs:01}. In
particular, with increasing optical depth of the silicate absorption
feature, the 8.6 \mum\, feature is decreased tremendously (see
  Fig. \ref{Prof6_ext}). To asses the influence of extinction on the
band profiles and their intensities, the "standard" extinction law
\citep{Draine:grain:85,Mathis:ext:90,Martin:ext:90} is applied to a
template PAH spectrum. This template spectrum is obtained by a
continuum divided spectrum of a source suffering no extinction. From
the resulting spectrum, the band profiles and band intensities were
derived in the same way as for the sources considered in this paper
(see Fig. \ref{Prof6_ext}).  Although the full PAH spectrum
changes significantly with increasing A$_K$, the normalised 6.2, 7.7
and 8.6 \mum\, band profiles derived with the above discussed
continuum determination are hardly affected by the applied extinction,
largely because the extinction is quite grey over this wavelength
region. But, the derived intensities and hence their ratios are
certainly influenced. For this work, no extinction
correction has been applied.

Five sources show water ice absorption at 6.0 \mum \,(see
Table \ref{sample}) and hence the profile of the 6.2 \mum\, feature
can be influenced. In view of the profile of the water ice band, its
influence would be expected to be strongest on the blue side of the 6.2 \mum\,
feature. However, the ice absorption is very small in our
sources and the 6.2 \mum\, band is situated in the red non-steep
wing of the water band. Therefore,  its influence on the 6.2
\mum \, band profile (FWHM and peak position) is negligible
\citep[see][]{Spoon:silenpahs:01}.

\subsubsection{Normalisation of the profiles}

\begin{figure}[!th]
  \begin{center}
   \begin{picture}(240,240)(-5,10)
   \psfig{figure=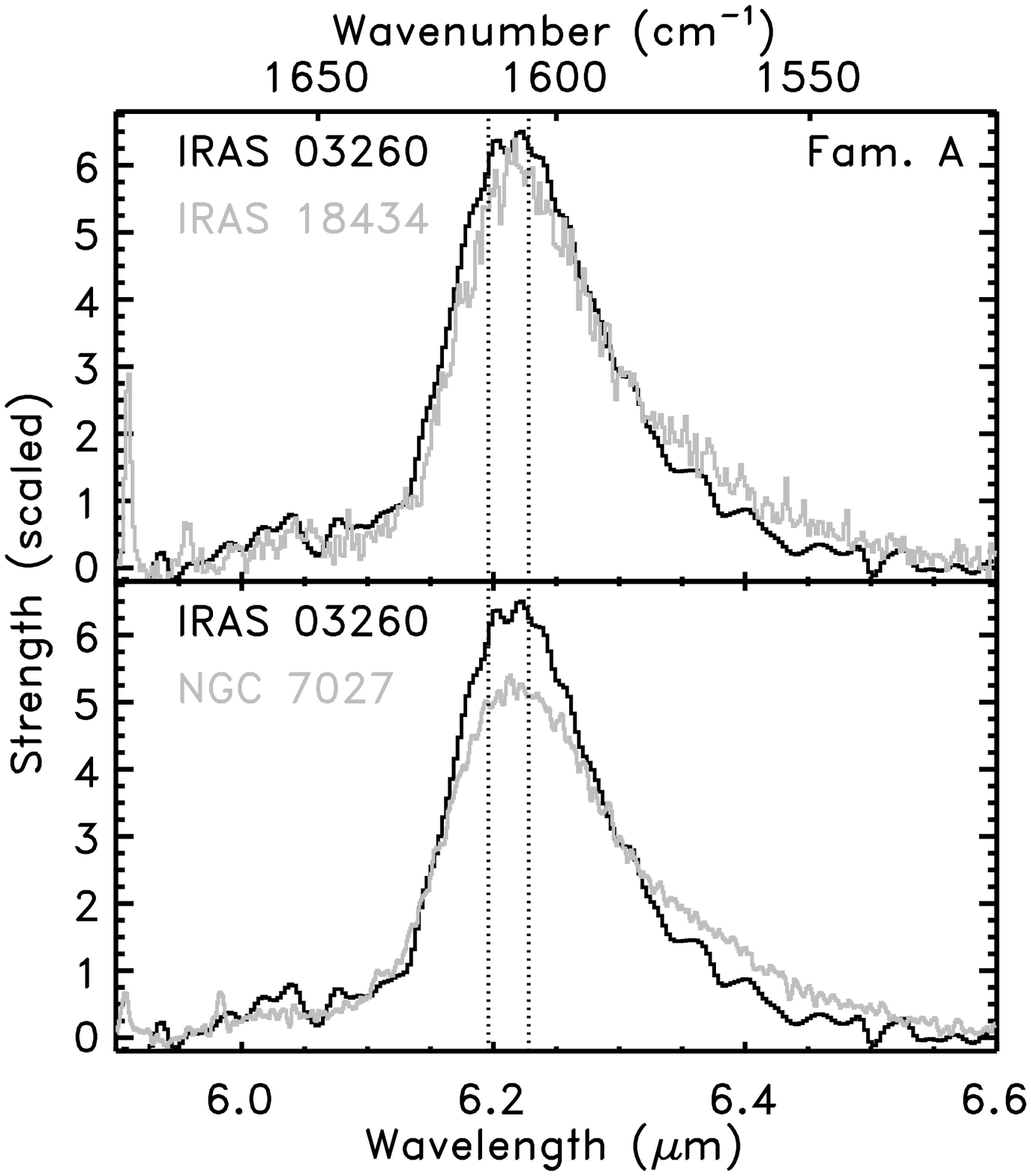,width=240pt,height=240pt}
   \end{picture}
  \end{center}
\caption{The normalised PAH CC stretching features of class A.  The spectra
  shown exemplify the variations inherent in this class. The
  vertical dotted lines show the range in peak positions in class A. }
\label{varA}
\end{figure}

For comparison, the 6.2 \mum\, profiles shown in Figs. \ref{varA},
\ref{var62}, \ref{varB1}, \ref{varC}, \ref{fit} and \ref{gauss} are
scaled in such a way that the integrated flux within the profile is
equal to 1.  In this normalization procedure, the lower limit
(6.1 \mum) is chosen so as to exclude the 6.0 \mum\, emission feature,
while the upper limit is determined by the most extreme end of the
feature (i.e. 6.6 \mum).  Analogously, the 7.7 \mum\,band profiles
shown in Figs.  \ref{varA77}, \ref{varC77}, \ref{varie77B} and
\ref{gauss7678} were normalised so that the total flux from 7.2 to 8.2
\mum\, equals one. For the sources IRAS~17347, IRAS~07027, He2-113 and
HD~100546, we normalised the spectra so that the total flux from
7.2 to 8.35 \mum\, equals one in order to cover the 7.7 \mum\, complex
completely.  The 8.6 \mum\, band profiles were normalised so that the
total flux from 8.2 to 8.9 equals one.

\section{The band profiles}

In this section, we show how the various types of PAH emission
spectra found in our sample can be rationalised into spectral
classes which correspond to different band profiles. This is done
independently for the three main features in the 6--9 \mum\, region
: the 6.2, 7.7 and 8.6 \mum\, bands. Note however that there is some
variability of the band shapes within a single class.

\subsection{The 6.2 $\mu$m feature}
\label{f62}

In this section, we classify the 6.2 \mum\, bands present in our
sample. In addition, a decomposition of the band profile into two
symmetric components is discussed.

\subsubsection{The profile of the 6.2 $\mu$m feature}

\begin{figure}[!th]
  \begin{center}
   \begin{picture}(240,310)(-5,10)
   \psfig{figure=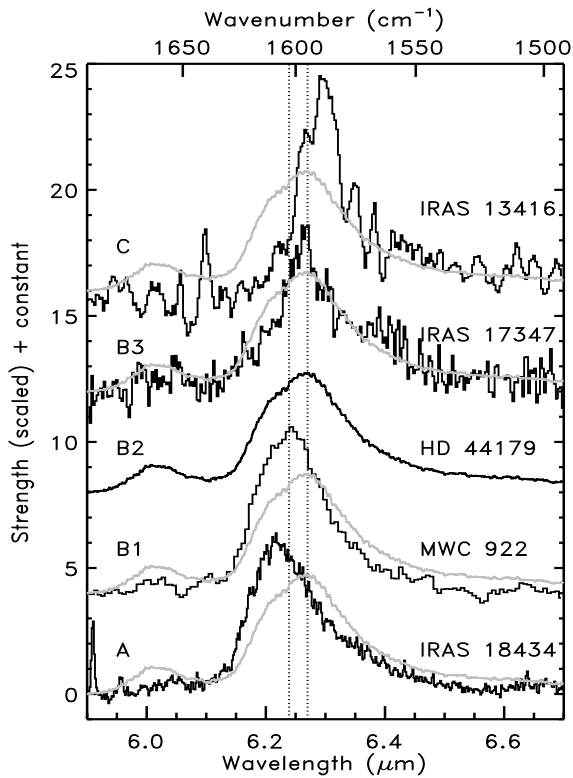,width=240pt,height=310pt,angle=90}
   \end{picture}
  \end{center}
\caption{The normalised PAH CC stretching features of class  B. For ease of comparison,
class A and C are represented by IRAS~18434 and IRAS~13416. As a
reference, the profile of HD~44179 is plotted on top of each
profile in grey. The vertical dotted lines show the range in peak positions
for class B. } 
\label{var62}
\end{figure}

The sources show a pronounced 6.2 \mum\, feature, sometimes preceded
by a weak feature at about 6.0 $\mu$m.  The emission profiles of the
6.2 \mum\, feature are distinctly asymmetric with a steep blue
rise and a red tail (see Fig. \ref{varA}).  Although the profile of
the 6.2 \mum\, feature is similar for all the sources, when
examined in detail significant differences become apparent.  A
definite range in peak positions is present in our sample, varying
between 6.19 and 6.29 $\mu$m (see Fig. \ref{var62}).  The width also
varies. The peak positions and FWHM values for all sources are given
in Table \ref{sample}.  Perusing the derived profiles, we recognise
three main classes, which we will designate by A, B, and C.

First, the majority of the 6.2 \mum\, bands peak between 6.19 and 6.23
$\mu$m.  This group will be referred to as class A. Note that
the strength of the red tail relative to the peak strength, and hence
the FWHM, varies within this class (see bottom panel of Fig.
\ref{varA}).  Furthermore, the top of the profile can be peaked or
rounded off (see top panel of Fig.  \ref{varA}).

\begin{figure}[!t]
  \begin{center}
   \begin{picture}(240,140)(-5,10)
   \psfig{figure=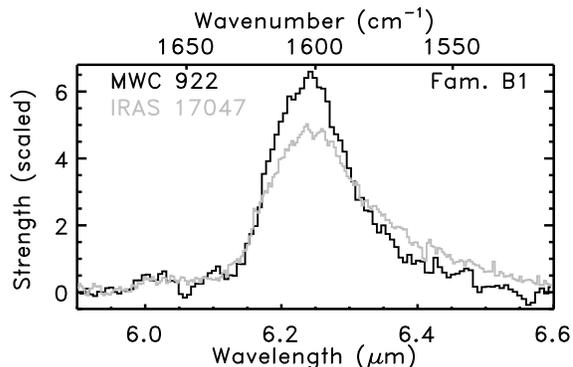,width=240pt,height=140pt}
   \end{picture}
  \end{center}
\caption{The normalised PAH CC stretching features of class B1.  These spectra show the
variations inherent in this grouping.}
\label{varB1}
\end{figure}

Second, the remaining sources have peak positions that range up to 6.29
$\mu$m.  We define members of class B as those having profiles with a peak
position between 6.235 and 6.28 \mum\, (see Table \ref{sample} and Fig. 
\ref{var62} and \ref{varB1}).  In general, the profiles of class B have a larger FWHM
compared to those of class A.  

\begin{figure}[!h]
  \begin{center}
   \begin{picture}(240,140)(-5,10)
   \psfig{figure=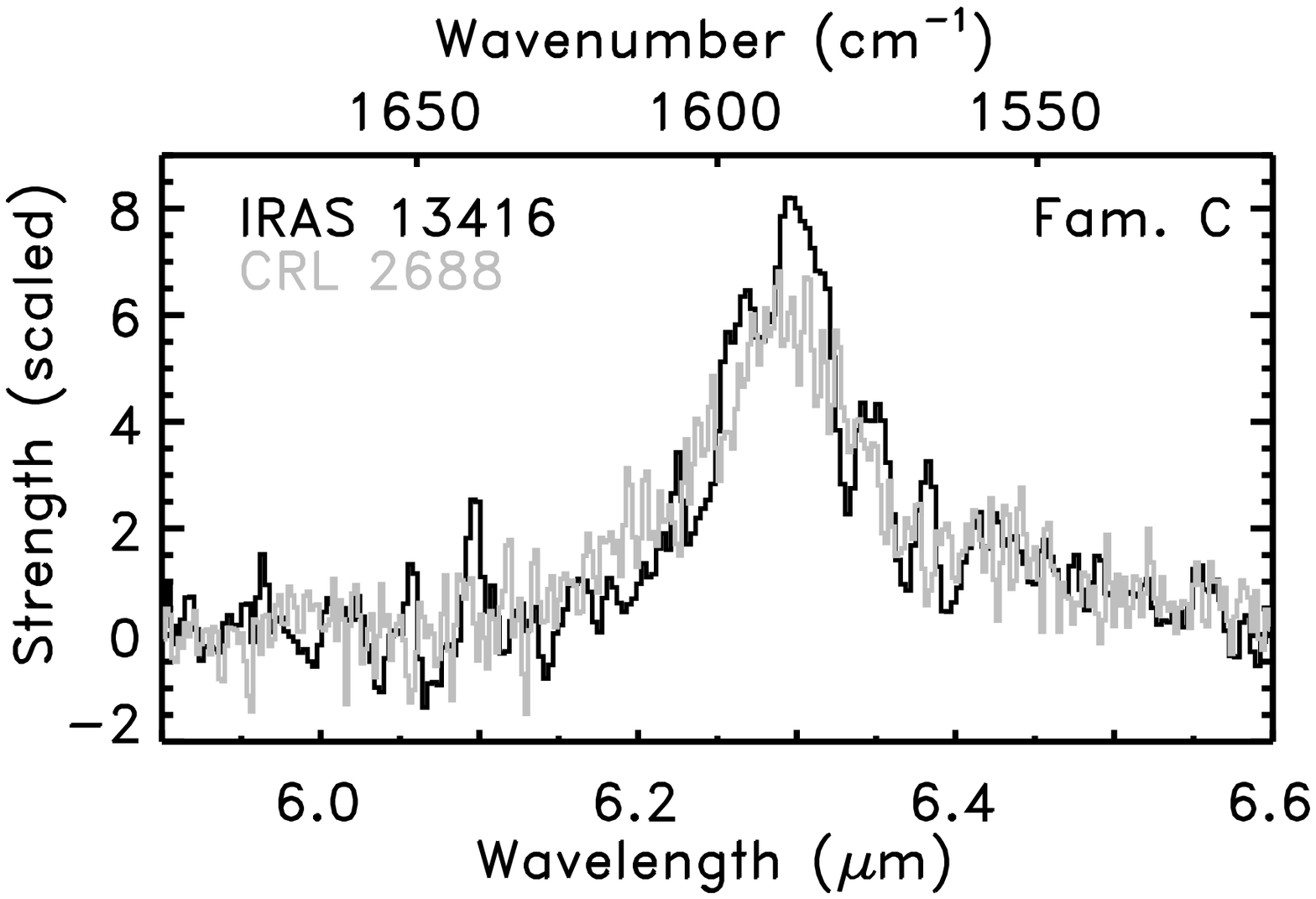,width=240pt,height=140pt}
   \end{picture}
  \end{center}
\caption{The normalised PAH CC stretching features of class  C.}
\label{varC}
\end{figure}

Class B can be further subdivided.  In particular, the peak position
of MWC~922 is shifted toward longer wavelengths compared to class A
but it has a similar profile.  Other members in this group (B1)
contain profiles which show a more pronounced red tail than MWC~922
(see Fig. \ref{varB1}).  These variations parallel those in class A.
Class B2 is represented by HD~44179 in Fig. \ref{var62}.  Their
profile is clearly less red-shaded compared to class B1 (and class A).
Note that the profile of HD~44179 shows substructure: one can
distinguish a blue shoulder.  As can be seen from Fig. \ref{varB1}, the
first 'component' occurs at the peak position of class A.  This is the
only source where substructure is revealed so clearly. In contrast to
the previous classes (B1 and A), no variations occur in the relative
strength of the red tail within this group. Class B3 (exemplified by
IRAS~17347 in Fig.  \ref{var62}) clearly lacks emission before 6.23
\mum\, and has a somewhat higher peak-to-red-tail ratio than B2
sources.  This subdivision of group B is somewhat arbitrary.
Examining Fig.  \ref{var62}, there seems to be a gradual progression
of the peak position to the red from bottom to top.

Third, IRAS~13416 shows a different profile peaking at 6.29 $\mu$m. We
classify objects with this last profile as belonging to class C.
Although the emission band in this source is very weak and noisy, it
is clear that the profile is more symmetric than in other
  classes and its FWHM is the smallest found in our sample.  The '6.2
\mum' band in CRL~2688 peaks at 6.29 \mum\, and exhibits the same
profile as IRAS~13416.  Hence, this source is classified as a member
of class C (see Fig.  \ref{varC}).

\begin{table*}[!th]
\caption{Spectral classification of sources, description of band
    profiles and band strength ratios.}
\label{sample}
  \begin{center}
    \leavevmode
    \scriptsize
{\setlength{\tabcolsep}{4pt}
\begin{tabular}{lccccccccccc}
\hline\\[-0.2cm]
Source          
& \multicolumn{4}{c}{6.2 \mum\,feature} 
& \multicolumn{2}{c}{7.7 \mum\,complex} 
& \multicolumn{2}{c}{8.6 \mum\,feature} 
& \multicolumn{1}{c}{I$_{7.6}$/I$_{7.8}$$^{c}$} 
& \multicolumn{1}{c}{I$_{7.6}$/I$_{6.2}$$^{c}$} & R$^{e}$\\
& \multicolumn{1}{c}{fam.}  
& \multicolumn{1}{c}{peak pos.} 
& \multicolumn{1}{c}{FWHM} 
& \multicolumn{1}{c}{I$_{2}$$^{a}$}
&\multicolumn{1}{c}{fam.}
& \multicolumn{1}{c}{peak pos.}    
&  \multicolumn{1}{c}{fam.}
& \multicolumn{1}{c}{peak pos.$^{b}$}&  & & \\
                &       & \multicolumn{1}{c}{[\mum ]}  & 
\multicolumn{1}{c}{[10$^{-2}$ \mum ]} 
& \multicolumn{1}{c}{\%} &  
& \multicolumn{1}{c}{[\mum ]} & 
&\multicolumn{1}{c}{[\mum ]} &  & &  \\[2.3pt] 
\hline\\[-0.2cm]           
NGC~253$^{\ddagger}$           & A    & 6.212$\pm$0.010 & 13.4$\pm$0.5 &41$\pm$12& A$'$  & 7.611$\pm$0.013 & A$''$ & 8.59& 1.02$\pm$0.14 & 0.89$\pm$0.13 & 0.58\\[0.7pt]
W~3A$^{\dagger}$               & A    & 6.223$\pm$0.017 & 11.1$\pm$0.3 &36$\pm$4 & A$'$  & 7.626$\pm$0.005 & A$''$ & 8.60& 1.84$\pm$0.37 & 1.50$\pm$0.21 & 0.34\\[0.7pt]
IRAS~02575$^{\dagger,\sharp}$  & A    & 6.227$\pm$0.008 & 13.3$\pm$2.2 &39$\pm$2 & A$'$  & 7.696$\pm$0.108 & A$''$ & 8.60& 0.85$\pm$0.12 & 0.96$\pm$0.14 & 0.34\\[0.7pt]
IRAS~03260                     & A    & 6.216$\pm$0.010 & 13.2$\pm$0.2 &34$\pm$2 & A$'$  & 7.622$\pm$0.027 & A$''$ & 8.60& 1.56$\pm$0.22 & 0.98$\pm$0.14 & 0.78\\[0.7pt]
Orion PK1$^{\dagger , \sharp}$ & A    & 6.207$\pm$0.006 & 12.5$\pm$0.5 &41$\pm$2 & A$'$  & 7.637$\pm$0.015 & A$''$ & 8.60& 1.43$\pm$0.20 & 1.15$\pm$0.16 & 0.37\\[0.7pt]
Orion PK2$^{\sharp}$           & A    & 6.215$\pm$0.006 & 12.4$\pm$0.6 &36$\pm$2 & A$'$  & 7.632$\pm$0.006 & A$''$ & 8.60& 1.35$\pm$0.20 & 1.10$\pm$0.16 & 0.33\\[0.7pt]
Or.Bar~D8                      & A    & 6.206$\pm$0.017 & 15.2$\pm$0.2 &41$\pm$2 & A$'$  & 7.627$\pm$0.010 & A$''$ & 8.61& 1.32$\pm$0.42 & 0.93$\pm$0.20 & 0.56\\[0.7pt]
Or.Bar~BRGA                    & A    & 6.222$\pm$0.016 & 12.2$\pm$0.3 &39$\pm$3 & A$'$  & 7.635$\pm$0.007 & A$''$ & 8.62& 1.32$\pm$0.19 & 0.93$\pm$0.13 & 0.67\\[0.7pt]
Or.Bar~D5                      & A    & 6.225$\pm$0.006 & 12.4$\pm$0.2 &44$\pm$2 & A$'$  & 7.634$\pm$0.007 & A$''$ & 8.61& 1.35$\pm$0.38 & 0.93$\pm$0.18 & 0.80\\[0.7pt]
Or.Bar~H2S1                    & A    & 6.212$\pm$0.011 & 12.2$\pm$0.1 &41$\pm$2 & A$'$  & 7.627$\pm$0.010 & A$''$ & 8.61& 1.47$\pm$0.29 & 0.94$\pm$0.13 & 0.70\\[0.7pt]
Or.Bar~D2                      & A    & 6.217$\pm$0.016 & 11.7$\pm$1.6 &44$\pm$10& A$'$  & 7.619$\pm$0.027 & A$''$ & 8.62& 1.15$\pm$0.34 & 0.73$\pm$0.14 & 0.52\\[0.7pt]
NGC~2023                       & A    & 6.214$\pm$0.010 & 11.0$\pm$1.0 &39$\pm$5 & A$'$  & 7.609$\pm$0.069 & A$''$ & 8.59& 1.56$\pm$0.30 & 0.83$\pm$0.12 & 1.14\\[0.7pt]
HD~44179$^{\star ,\flat}$      & B2   & 6.268$\pm$0.007 & 17.6$\pm$0.4 &59$\pm$7 & B$'$  & 7.859$\pm$0.012 & B$''$ & 8.67& 0.42$\pm$0.10 & 0.44$\pm$0.10 & 0.28\\[0.7pt]
IRAS~07027                     & B2/3 & 6.268$\pm$0.006 & 16.0$\pm$1.4 &66$\pm$3 & B$'$  & 7.921$\pm$0.041 & B$''$ & 8.67& 0.30$\pm$0.13 & 0.36$\pm$0.15 & 0.30\\[0.7pt]
M~82$^{\dagger ?, \ddagger}$   & A    & 6.210$\pm$0.001 & 13.3$\pm$0.7 &44$\pm$2 & A$'$  & 7.626$\pm$0.008 & A$''$ & 8.61& 1.08$\pm$0.15 & 0.90$\pm$0.13 & 0.74\\[0.7pt]
HR~4049$^{\circ ,\flat}$       & B2   & 6.260$\pm$0.009 & 15.2$\pm$0.6 &61$\pm$3 & B$'$  & 7.869$\pm$0.037 & B$''$ & 8.67& 0.17$\pm$0.05 & 0.22$\pm$0.07 & 0.11\\[0.7pt]
IRAS~10589$^{\dagger?}$        & A    & 6.223$\pm$0.010 &  9.4$\pm$0.3 &33$\pm$8 & A$'$  & 7.630$\pm$0.008 & A$''$ & 8.62& 1.30$\pm$0.18 & 1.20$\pm$0.17 & 0.60\\  [0.7pt]
HD~97048                       & A    & 6.221$\pm$0.010 & 13.8$\pm$0.3 &34$\pm$14& AB$'$ & $\diamond$      & A$''$ & 8.62& 0.75$\pm$0.11 & 0.75$\pm$0.11 & 0.32\\[0.7pt]
HD~100546                      & B    & 6.251$\pm$0.022 & 13.1$\pm$1.0 &49$\pm$3 & B$'$  & 7.903$\pm$0.136 & B$''$ & 8.66& 0.36$\pm$0.10 & 0.41$\pm$0.11 & 0.27\\[0.7pt]
IRAS~12063$^{\dagger}$         & A    & 6.217$\pm$0.011 & 11.5$\pm$0.1 &42$\pm$2 & A$'$  & 7.626$\pm$0.024 & A$''$ & 8.59& 1.20$\pm$0.17 & 1.22$\pm$0.17 & 0.43\\ [0.7pt] 
IRAS~12073                     & A    & 6.205$\pm$0.005 & 10.7$\pm$1.6 &30$\pm$3 & A$'$  & 7.626$\pm$0.040 & d     & -   & 1.56$\pm$0.26 & 1.77$\pm$0.25 & 0.07\\[0.7pt]
IRAS~13416$^{\circ}$           & C    & 6.299$\pm$0.015 &  8.7$\pm$1.3 &100      & C$'$  & 8.199$\pm$0.059 & C$''$ & -   & -    &   -  & 0.21\\  [0.7pt]
circinus$^{\dagger, \ddagger}$ & A    & 6.210$\pm$0.014 & 10.0$\pm$2.5 &33$\pm$10& A$'$  & 7.616$\pm$0.014 & A$''$ & 8.60& 0.88$\pm$0.12&1.03$\pm$0.15 & 0.22\\[0.7pt]
HE~2-113                       & B2   & 6.255$\pm$0.007 & 16.7$\pm$0.4 &64$\pm$9 & B$'$  & 7.913$\pm$0.043 & B$''$ & 8.63& 0.35$\pm$0.13 & 0.31$\pm$0.12 & 0.29\\[0.7pt]
IRAS~15384$^{\dagger}$         & A    & 6.222$\pm$0.005 & 13.6$\pm$0.1 &41$\pm$3 & A$'$  & 7.618$\pm$0.015 & A$''$ & 8.61& 1.40$\pm$0.22 & 1.21$\pm$0.17 & 0.71\\[0.7pt]
G~327$^{\dagger}$              & A/B1 & 6.228$\pm$0.025 & 14.6$\pm$1.0 &41$\pm$4 & A$'$  & 7.619$\pm$0.028 & A$''$ & 8.59& 1.31$\pm$0.19 & 1.14$\pm$0.16 & 0.70\\[0.7pt]
IRAS~15502$^{\dagger}$         & A    & 6.211$\pm$0.008 & 10.2$\pm$0.8 &39$\pm$2 & A$'$  & 7.589$\pm$0.027 & d     & -   & 1.50$\pm$0.22 & 1.14$\pm$0.16 & 0.54\\[0.7pt]
IRAS~16279                     & A    & 6.219$\pm$0.017 & 19.0$\pm$0.3 &43$\pm$2 & A$'$  & 7.633$\pm$0.031 & A$''$ & 8.60& 0.85$\pm$0.13 & 0.77$\pm$0.11 & 0.33\\  [0.7pt]
IRAS~16594                     & A    & 6.227$\pm$0.029 & 14.3$\pm$2.2 &39$\pm$9 & A$'$  & 7.621$\pm$0.035 & A$''$ & 8.59& 1.32$\pm$0.20 & 0.65$\pm$0.09 & 0.26\\[0.7pt]
CD~-42 11721(off)$^{\dagger?}$ & A    & 6.224$\pm$0.008 & 12.5$\pm$0.4 &39$\pm$6 & A$'$  & 7.612$\pm$0.020 & A$''$ & 8.60& 1.51$\pm$0.21 & 0.93$\pm$0.13 & 0.84\\[0.7pt]
CD~-42 11721$^{\dagger?}$      & A    & 6.212$\pm$0.014 & 13.0$\pm$0.2 &36$\pm$2 & A$'$  & 7.609$\pm$0.016 & A$''$ & 8.60& 1.63$\pm$0.23 & 1.12$\pm$0.16 & 0.26\\[0.7pt]
IRAS~17047$^{\star ,\flat}$    & B1   & 6.246$\pm$0.012 & 17.1$\pm$0.8 &54$\pm$3 & B$'$  & 7.830$\pm$0.026 & B$''$ & 8.64& 0.52$\pm$0.10 & 0.42$\pm$0.08 & 0.16\\[0.7pt]
IRAS~17279$^{\dagger}$         & A    & 6.215$\pm$0.005 & 12.2$\pm$0.4 &30$\pm$2 & A$'$  & 7.622$\pm$0.023 & A$''$ & 8.60& 1.12$\pm$0.16 & 0.89$\pm$0.13 & 0.89\\ [0.7pt]
IRAS~17347                     & B3   & 6.259$\pm$0.011 & 11.0$\pm$4.7 &64$\pm$8 & B$'$  & 7.972$\pm$0.033 & B$''$ & 8.70& 0.16$\pm$0.08 & 0.25$\pm$0.12 & 0.45\\[0.7pt]
XX-OPH$^{\circ}$               & B3   & 6.270$\pm$0.027 & 14.5$\pm$4.7 &62$\pm$3 & B$'$  & 7.848$\pm$0.052 & B$''$ & 8.66& 0.27$\pm$0.07 & 0.23$\pm$0.06 & 0.11\\[0.7pt]
Hb 5$^{\triangleleft}$         & A    & 6.223$\pm$0.029 & 13.0$\pm$2.7 &34$\pm$2 & -     & -               & A$''$ & 8.61& 0.90$\pm$0.13 & 0.62$\pm$0.09 & 0.39\\[0.7pt]
IRAS~18032$^{\dagger}$         & A    & 6.209$\pm$0.016 & 11.8$\pm$0.2 &34$\pm$3 & A$'$  & 7.613$\pm$0.016 & A$''$ & 8.60& 1.19$\pm$0.17 & 1.25$\pm$0.18 & 0.82\\ [0.7pt]
IRAS~18116$^{\dagger ?}$       & A    & 6.227$\pm$0.007 & 11.3$\pm$1.4 &41$\pm$4 & A$'$  & 7.630$\pm$0.009 & A$''$ & 8.60& 1.17$\pm$0.17 & 0.88$\pm$0.12 & 0.73\\ [0.7pt]
GGD-27 ILL$^{\dagger, \sharp}$ & A    & 6.205$\pm$0.009 & 13.1$\pm$0.8 &38$\pm$6 & A$'$  & 7.603$\pm$0.010 & A$''$ & 8.60& 1.15$\pm$0.16 & 1.51$\pm$0.21 & 0.57\\[0.7pt]
MWC~922$^{\flat}$              & B1   & 6.243$\pm$0.006 & 13.3$\pm$0.2 &44$\pm$8 & A$'$  & 7.665$\pm$0.030 & $^{\natural}$& 8.62& 0.73$\pm$0.10 & 0.67$\pm$0.09 & 0.17\\[0.7pt]
IRAS~18317$^{\dagger}$         & A    & 6.224$\pm$0.006 & 13.3$\pm$0.6 &41$\pm$2 & A$'$  & 7.634$\pm$0.007 & A$''$ & 8.59& 1.29$\pm$0.19 & 1.04$\pm$0.15 & 0.62\\[0.7pt]
IRAS~18434$^{\dagger}$         & A    & 6.215$\pm$0.003 & 12.1$\pm$0.4 &43$\pm$2 & A$'$  & 7.637$\pm$0.010 & $^{\natural}$& - & 1.51$\pm$0.21 & 0.77$\pm$0.11 & 0.34\\[0.7pt]
IRAS~18502$^{\dagger}$         & A    & 6.214$\pm$0.022 & 16.0$\pm$0.6 &44$\pm$5 & A$'$  & 7.623$\pm$0.020 & A$''$ & 8.59& 1.02$\pm$0.15 & 1.07$\pm$0.15 & 0.85\\     [0.7pt]
HD~179218                      & B1/2 & 6.257$\pm$0.032 & 18.6$\pm$3.0 &54$\pm$3 & B$'$  & 7.786$\pm$0.126 & B$''$ & 8.65& 0.27$\pm$0.04 & 0.25$\pm$0.04 & 0.20\\[0.7pt]
IRAS~18576$^{\circ}$           & B1   & 6.249$\pm$0.027 & 12.2$\pm$0.3 &47$\pm$8 & A$'$  & 7.653$\pm$0.084 & A$''$ & 8.58& 0.86$\pm$0.12 & 0.74$\pm$0.10 & 0.59\\[0.7pt]
BD~+30 3639$^{\flat}$          & B1   & 6.239$\pm$0.008 & 14.0$\pm$0.6 &46$\pm$3 & B$'$  & 7.842$\pm$0.009 & B$''$ & 8.64& 0.34$\pm$0.05 & 0.38$\pm$0.05 & 0.32\\[0.7pt]
IRAS~19442$^{\dagger, \sharp ?}$ & A  & 6.218$\pm$0.008 & 13.6$\pm$0.3 &44$\pm$2 & A$'$  & 7.614$\pm$0.019 & A$''$ & 8.61& 1.30$\pm$0.18 & 1.04$\pm$0.15 & 0.64\\[0.7pt]
BD~+40 4124                    & A    & 6.203$\pm$0.011 &  9.8$\pm$0.5 &20$\pm$1 & A$'$  & 7.603$\pm$0.034 & A$''$ & 8.60& 1.80$\pm$0.31 & 1.13$\pm$0.16 & 0.16\\[0.7pt]
S~106 (IRS4)$^{\dagger}$       & A    & 6.210$\pm$0.014 & 13.9$\pm$0.3 &41$\pm$4 & A$'$  & 7.621$\pm$0.016 & A$''$ & 8.61& 1.43$\pm$0.20 & 1.05$\pm$0.15 & 0.30\\ [0.7pt]
NGC~7023I                      & A    & 6.213$\pm$0.013 & 13.4$\pm$1.4 &41$\pm$10& A$'$  & 7.598$\pm$0.022 & A$''$ & 8.60& 1.88$\pm$0.35 & 0.83$\pm$0.12 & 1.11\\[0.7pt]
CRL~2688$^{\flat}$             & C    & 6.290$\pm$0.015 & 11.1$\pm$1.2 &98$\pm$5 & C$'$  & 8.202$\pm$0.062 & C$''$ & -   & -    &   -  & 0.12\\ [0.7pt]
NGC~7027$^{\star ,\flat}$      & A    & 6.213$\pm$0.002 & 15.6$\pm$0.2 &44$\pm$3 & B$'$  & 7.814$\pm$0.001 & B$''$ & 8.64& 0.64$\pm$0.09 & 0.42$\pm$0.06 & 0.33\\   [0.7pt]
IRAS~21190                     & A    & 6.210$\pm$0.001 & 11.6$\pm$1.6 &36$\pm$3 & A$'$  & 7.612$\pm$0.024 & A$''$ & 8.61& 1.56$\pm$0.23 & 0.77$\pm$0.11 & 0.41\\  [0.7pt]
IRAS~21282                     & A    & 6.213$\pm$0.005 & 14.6$\pm$0.2 &44$\pm$5 & AB$'$ & $\diamond$      & A$''$ & 8.62& 0.64$\pm$0.09 & 0.61$\pm$0.09 & 0.41\\[0.7pt]
IRAS~22308                     & A    & 6.205$\pm$0.010 & 13.4$\pm$0.2 &37$\pm$5 & A$'$  & 7.614$\pm$0.023 & A$''$ & 8.61& 1.40$\pm$0.20 & 1.12$\pm$0.16 & 0.93\\ [0.7pt]
IRAS~23030                     & A    & 6.205$\pm$0.014 & 11.0$\pm$2.1 &32$\pm$6 & A$'$  & 7.625$\pm$0.054 & A$''$ & 8.60& 1.21$\pm$0.20 & 1.02$\pm$0.14 & 0.55\\ [0.7pt]
IRAS~23133                     & A    & 6.217$\pm$0.004 & 11.9$\pm$0.5 &39$\pm$2 & A$'$  & 7.631$\pm$0.011 & A$''$ & 8.60& 1.20$\pm$0.17 & 0.93$\pm$0.13 & 0.84\\[1pt]
\hline\\[-0.5cm]         
\end{tabular}}
  \end{center}
\small
$^a$: intensity fraction of comp. 2 (see Sect. \ref{decompose});  $\;$ 
$^b$: peak position derived by fitting a Gaussian to the profile. The
error on the position is dominated by the error on the continuum
and is $<$0.015;  $\;$ 
$^c$: based upon the Gaussian fit to the 7.7 \mum \, complex (see
Sect. \ref{f77}); $\;$ 
$^e$: $\Sigma$PAH/cont [6-9 \mum] (see Sect. \ref{corr})$\;$  
d: feature detected; $\;$  
$^\triangleleft$: strong [NeVI] present on top of the 7.6 \mum \,
feature,  and hampers the determination of the position;   $\;$ 
$^\diamond$: the peak intensity of the 7.6 and 7.8 \mum \, features are
equal within the estimated error; $\;$   
$^\natural$: sources with an unusual 8.6
\mum\, feature (see Sect. \ref{f86});   $\;$ 
$^\star$ and $^\flat$: sources suffering from memory effects at
resp. 6 and 7-9 \mum;  $\;$  
$^\sharp$: water ice absorption (6 \mum) present; $\;$  
$^\dagger$: silicate absorption (9.7 \mum) present; $\;$ 
$^\circ$: freedom in cont. determination at 6 \mum; $\;$
$^\ddagger$: redshifted corrected. Redshifts are taken from the NASA/IPAC
Extragalactic Database.\\
\end{table*}

\begin{figure}[!th]
  \begin{center}
   \begin{picture}(200,320)(-5,10)
   \psfig{figure=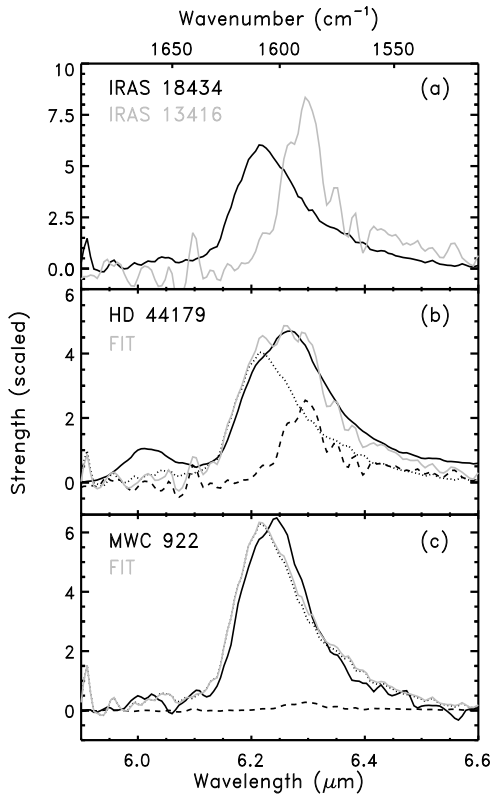,width=200pt,height=320pt}
   \end{picture}
  \end{center}
\caption{Panel a shows the two components used for fitting the normalised 
  profiles in class B. Panel b and c show the normalised profiles of
  He~2-113 and MWC~922 respectively together with the fit and the two
  components, reflecting their respective contributions. The spectra
  are interpolated to the same wavelength grid of resolution 500.}
\label{fit}
\end{figure}

We did not detect in our sample the 6.35 \mum\, band reported by 
\citet{Verstraete:m17:96} in the HII region of M17-SW. In fact, a
closer inspection of this data suggests that the 6.35 \mum\, feature
is the result of a cosmic ray hit on the SWS detectors. Therefore,
we do not include the 6.35 \mum\, band in our analysis.

Variations are also seen in the strength of the weak 6.0 \mum\,
feature. It seems to be uncorrelated with the 6.2 \mum\, band in both
strength and peak position (contrast e.g. HD~44179 and He~2-113,
cf. Figs \ref{var62} and \ref{fit}). 
  
\subsubsection{Decomposing the 6.2 \mum\, feature}
\label{decompose}
 
\begin{figure}[!thb]
  \begin{center}
   \begin{picture}(240,140)(-5,10)
   \psfig{figure=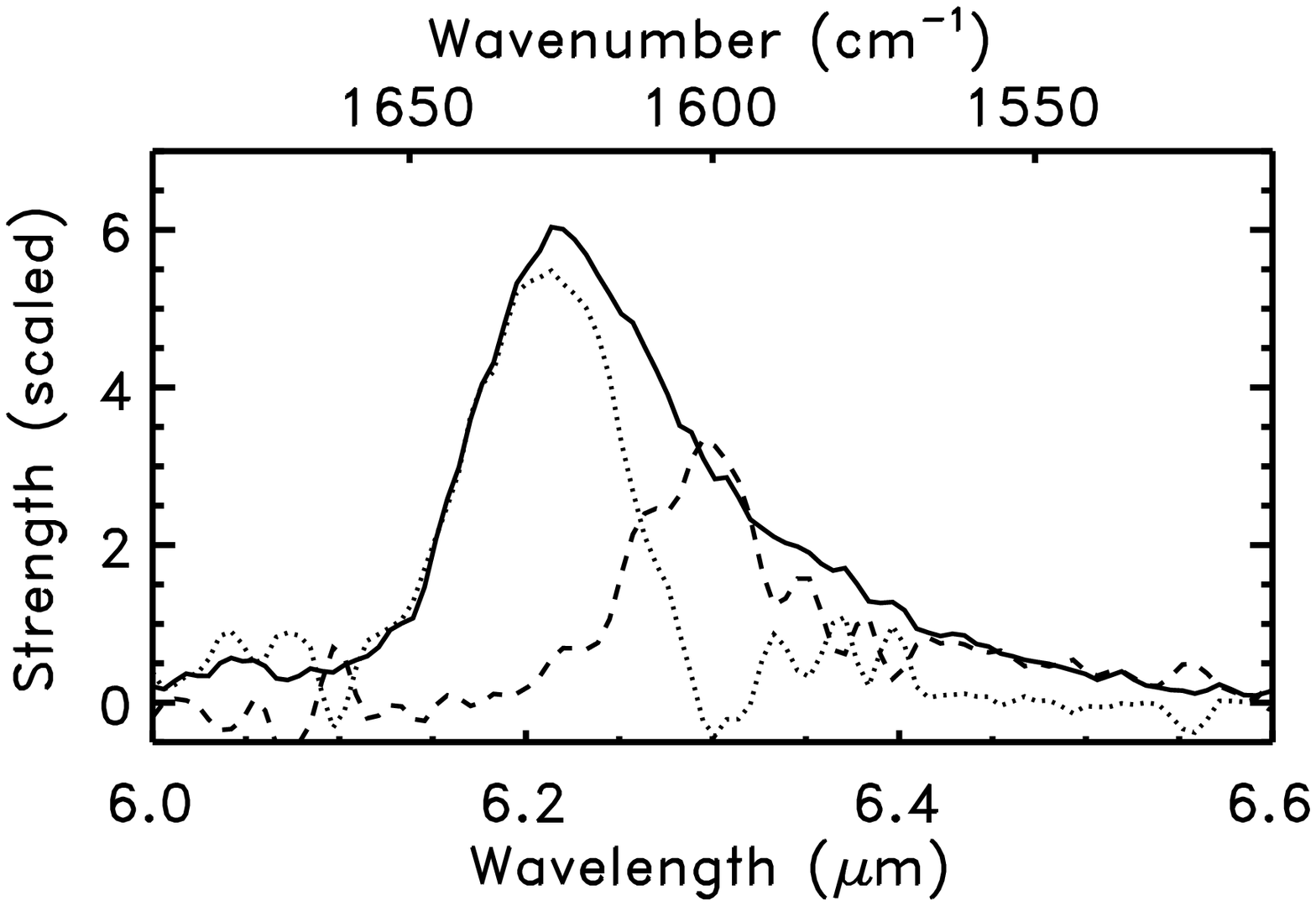,width=240pt,height=140pt}
   \end{picture}
  \end{center}
\caption{The normalised 6.2 \mum\, profile of IRAS~18434 (solid line) with the
  possible fraction of component 2, i.e. class C (dashed line). The
  dotted line represents the Gaussian profile derived through
  subtraction.  The spectra are interpolated to the same wavelength
  grid of resolution 500.}
\label{gauss}
\end{figure}

Because the peak position seems to gradually and smoothly increase, we
hypothesise that all profiles of class B consist of a combination of
two components with extreme peak position; one profile of class A and
one profile with the other extreme peak position at 6.29 $\mu$m, class
C.  For this two component analysis, we have adopted IRAS~18434 as
component 1 and IRAS~13416 as component 2.  In order to fit a
combination of these two components, we interpolated all spectra on
the same wavelength grid of resolution 500.  All profiles of class B
are in general well fit by this procedure (see Fig. \ref{fit}).
However, we note that for B2/3 spectra, the fit starts slightly
short-wards of the observed profile. The only exceptions are MWC~922
and IRAS~17347 which show some subtle differences (see Fig.
\ref{var62}). Longwards of 6.3 \mum, MWC~922 is fit well by component
1 and component 2 seems to be absent. However, the peak of MWC~922 is
shifted towards the red with respect to component 1. IRAS~17347
exhibits a similar profile as component 2 with slightly more emission
on the blue wing. However, the profile of IRAS~17347 is shifted to the
blue compared to component 2.

\begin{figure}[!t]
  \begin{center}
   \begin{picture}(220,220)(10,10)
   \psfig{figure=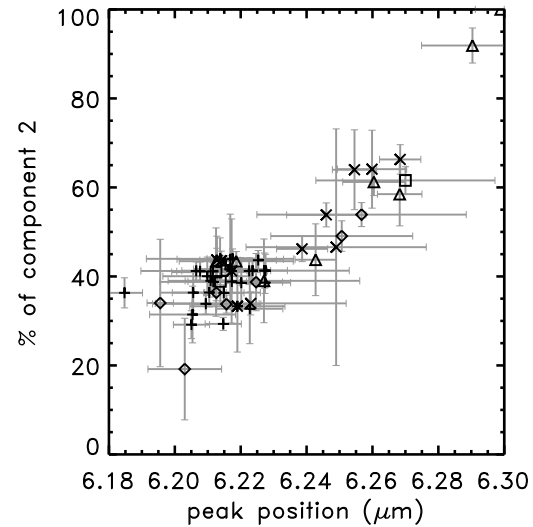,width=220pt,height=220pt}
   \end{picture}
  \end{center}
\caption{The peak position of the 6.2 \mum\, band plotted against 
  the fraction of component 2 in the profile (see text). The symbols
  are : HII regions +, Herbig AeBe stars $\diamond$, RNe/Be/sym
  $\Box$, Post-AGB stars $\triangle$, PNe $\times$ and galaxies $\ast$. }
\label{posvscompc}
\end{figure}

In fact, the component 1 profile we defined above (IRAS~18434)
may already include some contribution of component 2 (IRAS~13416).
To investigate this case, we have subtracted a scaled component 2
from the profile of IRAS~18434 (the adopted component 1); a 
symmetrical profile remains peaking at 6.21 and with a FWHM of 0.09
\mum\, (see Fig.  \ref{gauss}). Applying this method to all
  sources, the same symmetrical profile remains for all sources.
Hence, it is possible that the two intrinsic profiles are this derived
symmetric component and the intrinsically symmetric profile of
component 2. The fraction of component 2 varies from source to source
(see Table \ref{sample}) covering the full range from 20 to 100 \%.
The systematic uncertainty due to the imperfect template spectra can
be estimated by degrading the resolution of the template and the
fitted sources from 500 to 100 and is found to be 5\%. Based upon
independent analysis of the up and down scans, we have also
estimate the uncertainty in the fitting procedure associated with
statistical noise. In Table \ref{sample} we quote the statistical
uncertainty, unless it is less than this estimated systematic
uncertainty. Obviously, a strong correlation is present between the
fraction of component 2 and the peak position of the profile (see Fig.
\ref{posvscompc}).  The two sources whose detailed profile deviates
the most in this procedure (MWC~922 and IRAS~17347), still agree well
with the observed trend.  It is noteworthy that no source in our
sample shows this "derived" symmetric profile. One source, which we
excluded because of its strong ice and silicate absorption features,
does exhibit a 6.22 \mum\, profile which closely resembles this
derived symmetric profile. It will be discussed in a forthcoming paper
(Peeters et~al. 2002, in prep.). Furthermore, BD+40~4124 does
exhibit a symmetric profile but slightly redshifted with respect to
the profile derived through the present decomposition procedure.

\subsection{The 7.7 \mum \, band}
\label{f77}

\begin{figure}[!htb]
  \begin{center}
   \begin{picture}(220,240)(-5,10)
   \psfig{figure=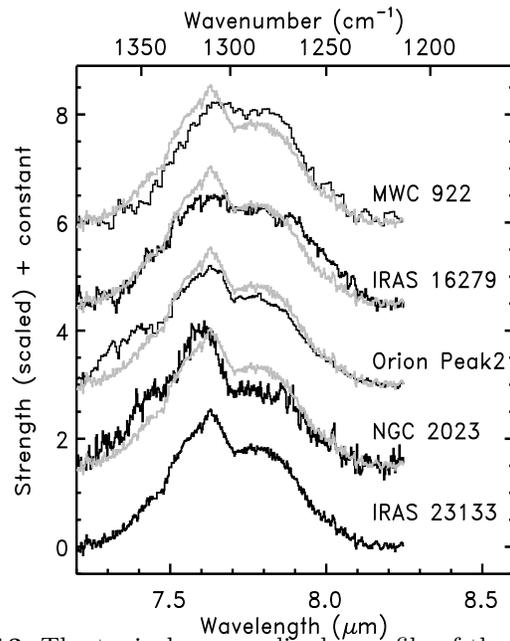,width=220pt,height=240pt}
   \end{picture}
  \end{center}
\caption{The typical - normalised - profile of the 7.7 \mum\, complex
  of class A$'$ represented by IRAS~23133, shown together with the
  particular cases within this class. As a reference, the typical 
    A$'$ profile of IRAS~23133 is plotted on top of each
  profile in grey.}
\label{varA77}
\end{figure}

Although the profile of the 7.7 \mum \, complex seems similar for most
sources, as with the 6.2 \mum\, band, significant differences become
apparent when this complex is examined in detail. Typically, this band
shows major subfeatures at $\sim$ 7.6 and $\sim$ 7.8 $\mu$m (cf. Figs.
\ref{prof6_fig_spectra} and \ref{varA77}) with possible minor
subfeatures near 7.3 to 7.4, 8.0 and 8.2 \mum. A definite range in
relative strength of the 7.6 versus 7.8 \mum \, component is present
in our sample, going from a dominant 7.6 \mum\, component toward a
dominant component peaking longwards of 7.7\mum.  Furthermore, the 7.7
\mum\, complex shifts as a whole. In particular, when the 7.6
component is not dominant, the peak position of the whole complex
varies from 7.79 to 7.97 \mum. Whether the peak position of the minor
7.6 component also varies in the latter case, cannot be
determined from the present dataset since it is then situated in
the wing of the dominant component.  In this paper, we will make a
distinction between those 7.7 \mum\, complexes with peak position
shortwards of 7.7 \mum, referred to as dominated by {\it the so-called
  7.6 \mum \, component} and those 7.7 \mum\, complexes with peak
position longwards of 7.7 \mum\, referred to as dominated by {\it the
  so-called 7.8 \mum\, feature}. This does not necessarily imply that
the apparent shift in peak position is due to a shift of the component
itself but can also be due to a different relative strengths of the
various components giving rise to the total 7.7 \mum\, complex. The
derived peak positions of the 7.7 \mum\, complex are given in Table
\ref{sample}.

\begin{figure}[!htb]
  \begin{center}
   \begin{picture}(220,140)(-5,10)
   \psfig{figure=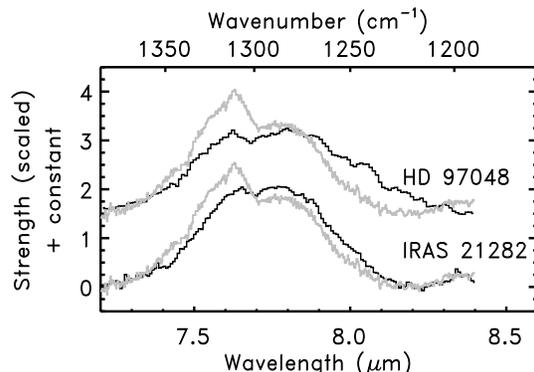,width=220pt,height=140pt}
   \end{picture}
  \end{center}
\caption{The normalised 7.7 \mum\, complex of class AB$'$. As a
  reference, the typical profile of class A$'$ (represented by
  IRAS~23133) is plotted on top of each profile in grey.} 
\label{varC77}
\end{figure}
 
\begin{figure}[!thb]
  \begin{center}
   \begin{picture}(240,400)(-5,10)
   \psfig{figure=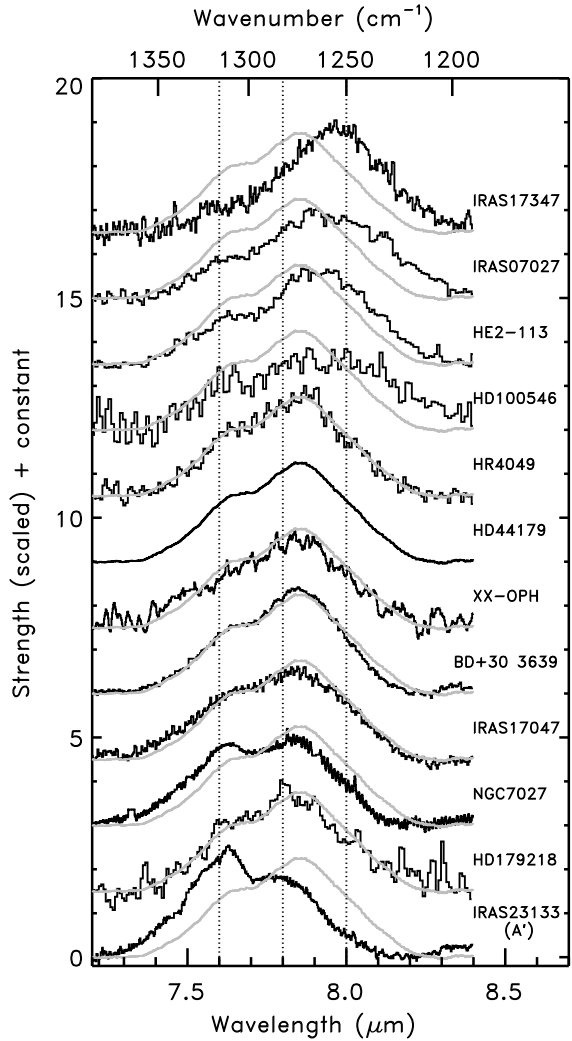,width=240pt,angle=90}
   \end{picture}
  \end{center}
\caption{The normalised 7.7 \mum\, complex of class B$'$. These spectra show
  the variations inherent in this grouping. As a reference, the bottom
  profile represents the typical profile of class A$'$ and the 7.7
  \mum \, profile of HD~44179 (a member of Class B$'$), is plotted on
  top of each profile in grey. The vertical dotted lines represent the
  nominal 7.6, 7.8 and 8 \mum\, positions. Clear shifts in peak
    position of the 7.7 \mum\, complex are present.}
\label{varie77B}
\end{figure}

Perusing the derived profiles, we recognise four main spectral
classes, which we will designate by A$'$, B$'$, AB$'$ and C$'$. This
classification is based upon the relative strength of the 7.6 and 7.8
\mum \, features. Sources where the 7.6 \mum \, feature is dominant,
group in class A$'$; class B$'$ contains sources which show a stronger
7.8 \mum\, feature; class AB$'$ groups those sources where the 7.6 and
7.8 have equal strength. Two sources do not show a 7.7 \mum \, complex
and hence form class C$'$.

When looking in detail at the normalised profiles of class A$'$, it is
clear that almost all sources have the same profile. Fig.
\ref{varA77} shows the possible variations within this class.
Orion Peak1, Orion Peak2, NGC~7023 and NGC~2023 show extra emission
between 7.2 and 7.4 \mum. NGC~2023 has also a more pronounced
I$_{p\, 7.6}$/I$_{p\, 7.8}$ ratio. IRAS~16279 shows extra
emission on the red wing compared to the typical profile of class
A$'$, indicating a contributor near 8.0 \mum. MWC~922 is
slightly redshifted as well. The latter two sources also show a less
pronounced ratio of the peak intensities of the 7.6 and 7.8 \mum \,
features.

Two sources, HD~97048 and IRAS~21282, have an equally strong 7.6 and
7.8 \mum \,feature (see Fig. \ref{varC77}) and form class AB$'$.
IRAS~21282 is also redshifted compared to the typical profile of class
A$'$, while HD~97048 has extra emission at the red wing near 8.0 \mum
\, compared to class A$'$.

Class B$'$ contains a collection of many different profiles (see Fig.
\ref{varie77B}). All of them are redshifted by different amounts
compared to the typical profile of class A$'$. The so-called 7.8 \mum
\, feature moves from 7.72 towards 7.97 \mum. The strength of the 7.6
\mum \, feature shows large variations (cf. NGC~7027 versus
IRAS~17347). Whether the 7.6 \mum \, feature behaves in a similar way
as the so-called 7.8 \mum \, feature concerning peak position, is
difficult to say since it is situated in the wing of the 7.8 \mum \,
feature.

Two sources, CRL~2688 and IRAS~13416, show no 7.7 \mum\, complex and
no 8.6 \mum\, feature (see \S\,\ref{f86}) but instead exhibit a
similar broad emission feature peaking at 8.22 \mum\, (see Fig.
\ref{varie77}). These two form group C$'$. \\ 

\begin{figure}[!tbh]
  \begin{center}
   \begin{picture}(220,275)(-5,10)
   \psfig{figure=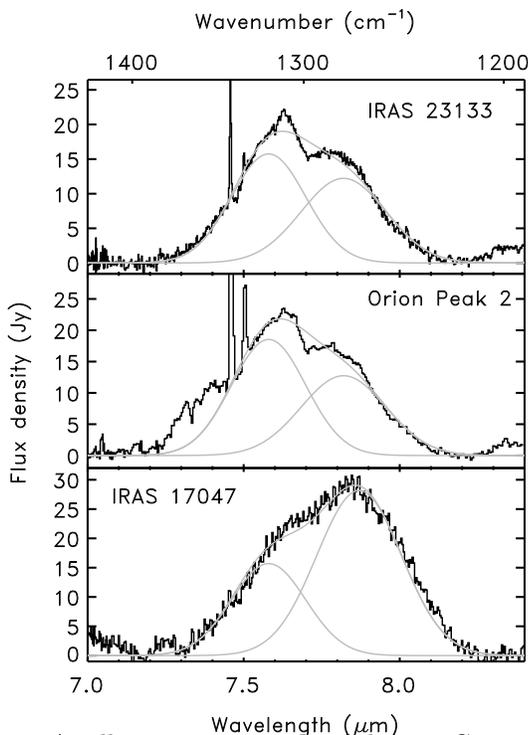,width=220pt,height=275pt}
   \end{picture}
  \end{center}
\caption{An illustrative example of the two Gaussians fitted to the 7.7 \mum
  \, complex.}
\label{gauss7678}
\end{figure}

A detailed fit to the profile can be obtained by fitting the total
region with, for example, several Lorentzian profiles
\citep{Verstraete:prof:01}.  Here we want to obtain an estimate
  of the relative flux contributions of the 2 main components rather
  than to fit the profile. In order to quantify these variations in
the relative strength of the 7.6 and the so-called 7.8 \mum \,
features, two Gaussians are fit to the 7.6 and the 7.8 \mum\, features
respectively. In a first attempt, the position and the width of
these Gaussians are fixed to $\lambda _{0}$(7.6)~=~7.58 \mum,
FWHM(7.6)~=~0.28 \mum\, and $\lambda _{0}$(7.8)~=~7.82 \mum,
FWHM(7.8)~=~0.32 \mum. For most sources, this works quite well (see
Fig. \ref{gauss7678}). The peak intensity of the 7.6 \mum\, band for
class A$'$ and the sharpness of the top of the 7.7 \mum\,
complex is however not well reproduced in all cases (see top panel of
Fig.  \ref{gauss7678}). With this method, the extra emission present
in four sources (Orion Peak 1, Orion Peak 2, NGC~2023 and NGC~7023) on
the blue side of the 7.7 \mum \, complex is ignored (see middle panel
of Fig. \ref{gauss7678}). In the cases of a few shifted 7.7 \mum\,
complexes, no good fit is obtained (i.e. for HD~100546, HD~44179,
He~2-113, BD~+30, IRAS~17047, IRAS~07027, IRAS~17347). For these
sources, a new fit is made in which the position of the 7.8 \mum\,
Gaussian is treated as a free parameter. The FWHM of the two Gaussians
is kept fixed although - for a single source - it might give a better
fit if the FWHM were a free parameter. However, to make a
consistent comparison between sources, we kept the bandwidths
constant. The relative strength of the 7.6 versus 7.8 \mum \,
feature as well as the intensity ratio of the 7.6 to the 6.2 \mum\,
feature are summarised in Table \ref{sample}. Class C$'$ is excluded
since it has no 7.7 \mum\, complex.  Based upon independent analysis
of the up and down scans, we estimate the uncertainty in the fitting
procedure associated with statistical noise to be generally less then
10\%. An estimate of the systematic uncertainty on the integrated
fluxes can be obtained by applying a different fit procedure and
comparing to our present 2-component decomposition. For that
reason, we fitted the 7.7 \mum\, complex with 4 Gaussians peaking at
7.5, 7.6, 7.8 and 8.0 \mum \, (Van Kerckhoven et~al. 2002, in prep.).
The strength of the combined 7.5 and 7.6 \mum\, Gaussians in this
method can be compared to that of the 7.6 \mum\, component applied
here.  Likewise, the combined 7.8 and 8.0 \mum\, Gaussians in this
method can be compared to the 7.8 \mum\, component applied here. This
4-component method provide better fits to the spectra with
obvious 7.5 \mum\, components (cf. Fig. \ref{gauss7678}, middle
panel). However, in general, the differences between the two methods
are small.  Hence, these intensity ratios are affected by systematic
and statistical uncertainties. In Table \ref{sample}, we quote the
larger of these two.

Sources dominated by the 7.6 \mum\, feature, always exhibit a 7.8
\mum\, feature. In contrast, sources with a dominant 7.8
\mum\, feature, do not always have a clear and distinctive 7.6 \mum\,
component. Especially, for sources where the 7.8 \mum \, feature peaks
at the longer wavelength end, the 7.6 \mum \, feature is extremely weak - if
present (e.g. IRAS~17347). Sources exhibiting the broad
emission feature at 8.22 \mum\, do not show a 7.7 \mum\, complex at
all (see Fig. \ref{varie77}).

\subsection{The 8.6 \mum \, feature}
\label{f86}

The 8.6 \mum \, profile derived with the chosen continuum is clearly
symmetric for all sources (see Fig. \ref{varie77}) and all of them
have the same FWHM.  Analogous to the 6.2 and 7.7 \mum, a definite
range in peak positions is present in our sample (see Fig.
\ref{varie77}). Perusing the derived profiles, we define 3 main
spectral classes A$''$, B$''$ and C$''$. Sources with peak position
ranging from 8.58 to 8.62 \mum\, group in class A$''$, sources with
peak position longwards of 8.62 \mum\, group in class B$''$. CRL~2688
does not exhibit a 8.6 \mum\, feature and forms group C$''$.
IRAS~13416 shows a weak feature near 8.6 \mum\, (see Fig.
\ref{varie77}). Careful examination of the independently reduced up
and down scans show that this feature is only present in one of the
two scans. Hence, we classify this source in group C$''$.  Table
\ref{sample} gives the class and the peak position derived by fitting
a Gaussian to the profile.  This classification relies heavily on the
adopted continuum.  When taking the general continuum (see Fig.
\ref{cont}, dashed line) instead of the local continuum (see Fig.
\ref{cont}, full line) to derive the profiles, the peak position of
the 8.6 \mum\, feature shifts for all sources slightly toward the
blue. As a result, the definition of the three classes changes
slightly. Two sources previously classified in class B$''$ now belong
to class A$''$ (i.e. NGC~7027, HD~100546). However, most sources
remain in their distinct classes.  Since the presence of a silicate
absorption feature has no major influence on the 8.6 \mum\, profile
(see Fig. \ref{Prof6_ext} and Sect. \ref{extinction}), the results
discussed here are valid for all sources.

MWC~922 shows an exceptionally strong 8.6 \mum \, feature. One HII
region, IRAS~18434, has a similarly strong 8.6 \mum \, feature.  The
latter source has a strong silicate absorption band but correction
for the silicate absorption will make this band even stronger (see
Fig. \ref{Prof6_ext}). A detailed study of these sources will be
presented in Hony et~al. (2002, in prep.) and Peeters et~al. (2002,
in prep.)  respectively.

\section{The classes}
\label{classes}

\begin{figure*}[!bht]
  \begin{center}
   \begin{picture}(540,240)(0,10)
   \psfig{figure=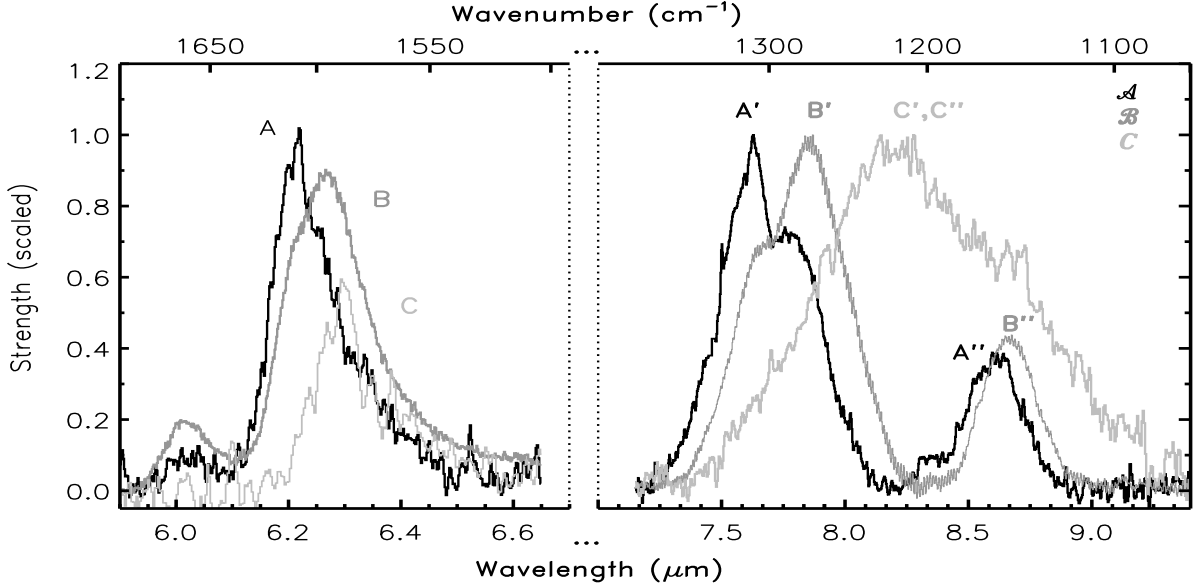,width=540pt,height=240pt,angle=90}
   \end{picture}
  \end{center}
\caption{An overview of the possible variations of the main PAH features in
  the 5-9 \mum \,region, i.e. the 6.2, 7.7 and 8.6 \mum\, features.
  The spectra are normalised so that the peak intensity of the
  strongest feature in the 8 \mum \,region equals one. Shown is (i)
  the CHII region IRAS~23133 illustrative of class \ca\, (i.e. A,
    A$'$ and A$''$) with a dominant 7.6 peak (black line), (ii)
  the Post-AGB star HD~44179 illustrative of class \cb\, (i.e. B,
    B$'$ and B$''$), peaking at $\sim$ 7.9 \mum\, (dark grey
    line) and (iii) the Post-AGB star IRAS~13416 representing class
  \cc\, (i.e. C, C$'$ and C$''$), with a broad emission feature
  at 8.22 \mum\, (light grey line). A clear shift in the 
  7.7 \mum\, complex is present between IRAS~23133 and HD~44179.}
\label{varie77}
\end{figure*}

In the previous sections, an independent study of the 6.2, 7.7 and 8.6
\mum\, features was made for our sample. For each of the features,
different classes were determined. Comparing those classes, an
interesting finding is made. The class classification of the different
bands correlate. This is illustrated in Fig. \ref{varie77}.  Sources
with a 6.2 \mum\, feature belonging to class A, have a 7.7 \mum\,
complex peaking at 7.6 \mum\, (class A$'$) together with a class A$''$
8.6 \mum\, feature and are referred to as class \ca\, sources; while
for those with a class B 6.2 \mum\, feature, the 7.7 \mum\, complex is
dominated by the so-called 7.8 \mum\, component (class B$'$) and their 8.6
\mum\, feature is shifted toward the red (class B$''$). The latter
sources are referred to as class \cb\, sources. The two sources
showing a single 6.3 \mum\, feature (class C) exhibit neither a 7.7
\mum\, complex nor an 8.6 \mum\, feature (class \cc).  Instead, both
sources show a broad emission feature at 8.22 $\mu$m.  The two sources
with an equally strong 7.6 and 7.8 \mum\, subfeature (i.e.  class
AB$'$), IRAS~21282 and HD~97048, exhibit a class A 6.2 \mum\, feature
and a class A$''$ 8.6 \mum\, feature. Note, however, that their 8.6
\mum\, feature peaks at the extreme end of class A$''$.  Possibly,
these two sources form an intermediate state between the spectrum
corresponding to class \ca\, and the spectrum corresponding to
class \cb. Another example of this type is NGC~7027.  Its 7.8 \mum\,
subfeature is slightly stronger in peak strength than its 7.6 \mum\,
subfeature, although by a small amount. However, this source belongs
to class A when considering the 6.2 \mum\, feature and class B$''$
concerning the 8.6 \mum\, feature.  Possibly, this is another
intermediate state.  IRAS~18576 and MWC~922 exhibit a B1 6.2 \mum\,
profile while their 7.7 and 8.6 \mum\, features both belong to classes
A$'$ and A$''$, respectively.  Fig.  \ref{varie77} provides an
overview of the PAH spectrum for each of the three main categories.
Although the classification of the 7.7 \mum\, complex was based on the
dominant subfeature, i.e. the 7.6 or 7.8 \mum\, subfeature, it is
clear from Fig. \ref{varie77} that the whole spectrum corresponding
with class \cb \, is shifted compared to that corresponding to class
\ca.

The two sources with the 6.3 \mum\, feature (IRAS~13416 and CRL~2688;
class C), show a similar spectrum composed of a rather weak 6.3 \mum\,
feature, a 8.22 \mum\, feature, an extremely weak 11.2 \mum\, feature
and a 3.3 \mum\, band (see Hony et~al. 2002, in prep.). The profiles
of the 8.22 \mum\, emission feature are identical, as are the 6.3
\mum\, profiles. The intensity ratio I$_{6.3}$/I$_{8.22}$ is however
not the same. Hence, the 6.3 and 8.22 \mum\,
features occur in similar conditions (object type, T, G$_{0}$, n$_{\rm
  e}$, etc.). But, based on these two sources,
it seems that no tight correlation, in terms of intensities, exist
between them. The 8.22 \mum\, feature in CRL~2688 and IRAS~13416 is
remarkably similar in peak position and width to the plateau
underneath the 7.7 and 8.6 \mum\, features subtracted from the
observed spectra (see Fig.\ref{cont} and Sect. \ref{continuum}). This
similarity is of course hard to prove. Here we do note that spatial
studies have shown that this plateau is an independent emission
component \citep[cf.][]{Bregman:hiivspn:89, Cohen:southerniras:89}.
Two sources which are not in our sample, IRAS~22272+5435 and
IRAS~07134+1004, both show a very broad feature at $\sim$ 8 \mum,
somewhat shortwards of the 8.22 \mum\, feature. This 8 \mum\, profile
is much broader then the 8.22 \mum\, feature and may be the result of
fortuitous convolution of a 7.7 \mum\, complex peaking at 7.6 \mum\,
and the 8.22 \mum\, feature.

\begin{table*}
\caption{ An overview of the classification of the 6.2, 7.7 and 8.6
  \mum\, features. The 6.2 and 8.6 \mum\, bands are classified by
  their peak position ($\lambda_{p}$) while the 7.7 \mum\, complex is
  classified by its dominant component, i.e. the 7.6 \mum \, component
  and/or the so-called 7.8 \mum\, component. See text for more details
  (Sects. \ref{f62}, \ref{f77} and \ref{f86}). For each spectral
  class, the range in the local radiation field G$_0$ and the spectral
  types within the class is given.}
\label{classdefinition}
\begin{center}
{\setlength{\tabcolsep}{0.2cm}
\begin{tabular}{clclclclcc}
\hline  \\[-5pt]
Class & \multicolumn{6}{c}{Characteristics} & \multicolumn{1}{c}{type
  of object} & range in G$_0$ & Sp. types\\
& \multicolumn{2}{c}{6.2 $\mu$m band} & \multicolumn{2}{c}{7.7 $\mu$m
  band} & \multicolumn{2}{c}{8.6 $\mu$m band}  & & & \\
 & & $\lambda_{p}$ & & comp. & & $\lambda_{p}$ & & & \\[5pt]
\hline \\[-5pt]
\ca & A   &     $\sim$ 6.22    &  A$'$   &  7.6   & A$''$ & $\sim$8.6
& HII, RN, galaxies, non-isolated & 3E2 to 7E6 & O,B\\
    &     &                    &         &        &       &
& Herbig Ae Be stars, PN : Hb5,       & & \\
    &     &                    &         &        &       &
& 2 Post-AGB stars : IRAS~16279,                   & & \\
    &     &                    &         &        &       &
&  IRAS~16594                    & & \\[8pt]
    &     &                    &  AB$'$  &  equal &       & 
& IRAS~21282, HD97048                       &  1.7E4 \&1E5 & O9 \&A0\\[8pt]
\cb & B   &     6.24--6.28     &  B$'$   &  "7.8" & B$''$ & $>$8.62
&isolated Herbig Ae Be stars, PNe,          &  6E4 to 2E7 & B,A,\\ 
&     &                    &         &           &       &
& 2 Post-AGB stars : HR4049, HD44179        &             & WC9-10\\[8pt]
\cc & C   &     $\sim$ 6.3     &  C$'$   &  8.22    & C$''$ & none
&2 Post-AGB stars : IRAS~13416,             & 5E3  & F5 \\
&     &                    &         &           &       &
& CRL~2688                                  &             & \\[5pt]
\hline
\end{tabular}}
\end{center}
\end{table*}

Considering the classification from an astronomical point of view,
 this analysis shows that the PAH spectrum in the 6 to 9 \mum\,
region, correlates with the type of source considered.  All HII
regions, reflection nebulae and all the extragalactic sources in this
sample have a class A 6.2 \mum\, feature, a 7.7 \mum\, complex peaking
at $\sim$ 7.6 \mum\, (class A$'$) and a class A$''$ 8.6 \mum \,
feature.  All Herbig AeBe stars that are still embedded in their
molecular cloud and have an HII region associated with them behave
like the ISM sources. The isolated Herbig AeBe stars, HD~179218 and
HD~100546, belongs to class \cb.  These two sources also show
crystalline silicates indicating that disk chemistry may well
influence the PAH population. The evolved stars are spread over the
different classes. Since their outflow is/was the place of birth of
the dust, they likely show the evolution of the PAH population in the
early phases of life. However, they do not reflect - at first sight -
a clear link between the PAH spectrum and the source evolutionary
state.  The two Post-AGB stars IRAS~13416 and CRL~2688, exhibit the
6.3 and 8.22 \mum\, features (class \cc). The two extremely metal-poor
sources HD~44179 and HR~4049 both exhibit a class \cb\,  spectrum
while the Post-AGB stars IRAS~16279 and IRAS~16594 display a class \ca\, spectrum. IRAS~21282 is similar to the latter two sources
except that it has an equally strong 7.6 and 7.8 \mum\, subfeature.
The PNe show similar variability in their spectra. Hb~5 is the only PN
in our sample showing a class \ca\, spectrum. IRAS~07027,
\mbox{He~2-113}, IRAS~17047, IRAS~17347 and BD~+30~3639 all show a
class \cb \, spectrum. In this class, sources with mixed PAH
classes are also present : IRAS~18576 and NGC~7027.

The PAH spectrum in the 6 to 9 \mum\, region apparently reflects local
physical conditions (object type, T, G$_{0}$, n$_{\rm e}$, etc.) or the
accumulated effect of processing from the formation sites in the AGB
or post-AGB phases to the ISM or the influence of disk chemistry.

\begin{figure*}[!thb]
  \begin{center}
   \begin{picture}(480,240)(0,0)
   \psfig{figure=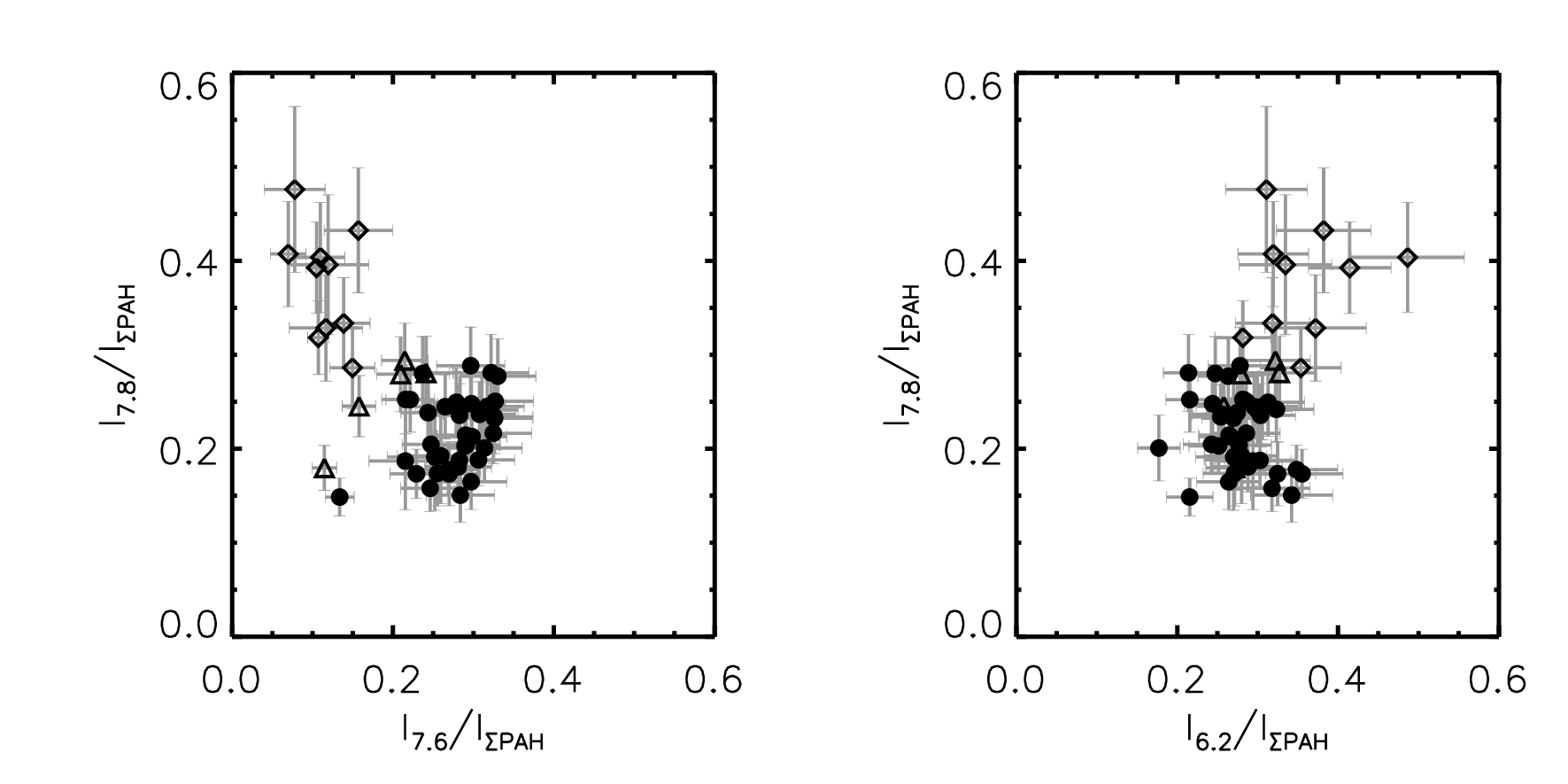,width=480pt,height=240pt}
   \end{picture}
  \end{center}
\caption{Shown is the integrated
  strength of the 7.8 \mum\, feature versus the 7.6 and 6.2 \mum\,
  features, normalised on the total flux emitted in the 3.3, 6.2, 7.7
  and 11.2 \mum\, features. Class \ca\, is represented by
  $\bullet$, class \cb\, by $\diamond$ and sources not belonging to
  these two classes by $\triangle$.}
\label{prof6_ratiosvssompah}
\end{figure*}

A detailed study on the variations in the PAH spectrum of
Herbig AeBe stars and the link with the local physical conditions will
be given in Van Kerckhoven et~al. (2002, in prep.). Hony et~al. (2002,
in prep.) concentrates on the PAH spectra of evolved stars and thus on the
evolution of the PAH population in the early phases while
Peeters et~al. (2002, in prep.) focuses on the HII regions.

\section{Correlation studies}
\label{corr}

Given the close relation between the main features in the 6 to 9
\mum\, region, it is of interest to investigate whether the strength
of these features can shed more light on the origin of the observed
relations and variations. Since the absolute intensities
are influenced by the intrinsic luminosity and distance of the source,
we study the variations in the relative strength of the PAH bands.
The intensity of the 7.6 and 7.8 \mum\, features
used here are integrated intensities of the Gaussian fits, while the
intensities of the other PAH features are integrated intensities of
the profile. The observed intensity ratios can be influenced by
extinction (see \S\,\ref{extinction}).  Since extinction strongly
influences the intensity of the 8.6 \mum\, feature, we do not
consider this feature here. 

When checking for connections between intensity ratios and object
type, we do detect a relation but there are always some exceptions
present. Nevertheless, when plotted against object class (classes \ca\,
and \cb), these exceptions disappear and clear segregations between
the two classes become apparent.

\begin{figure}[!bht]
  \begin{center}
   \begin{picture}(240,240)(0,10)
   \psfig{figure=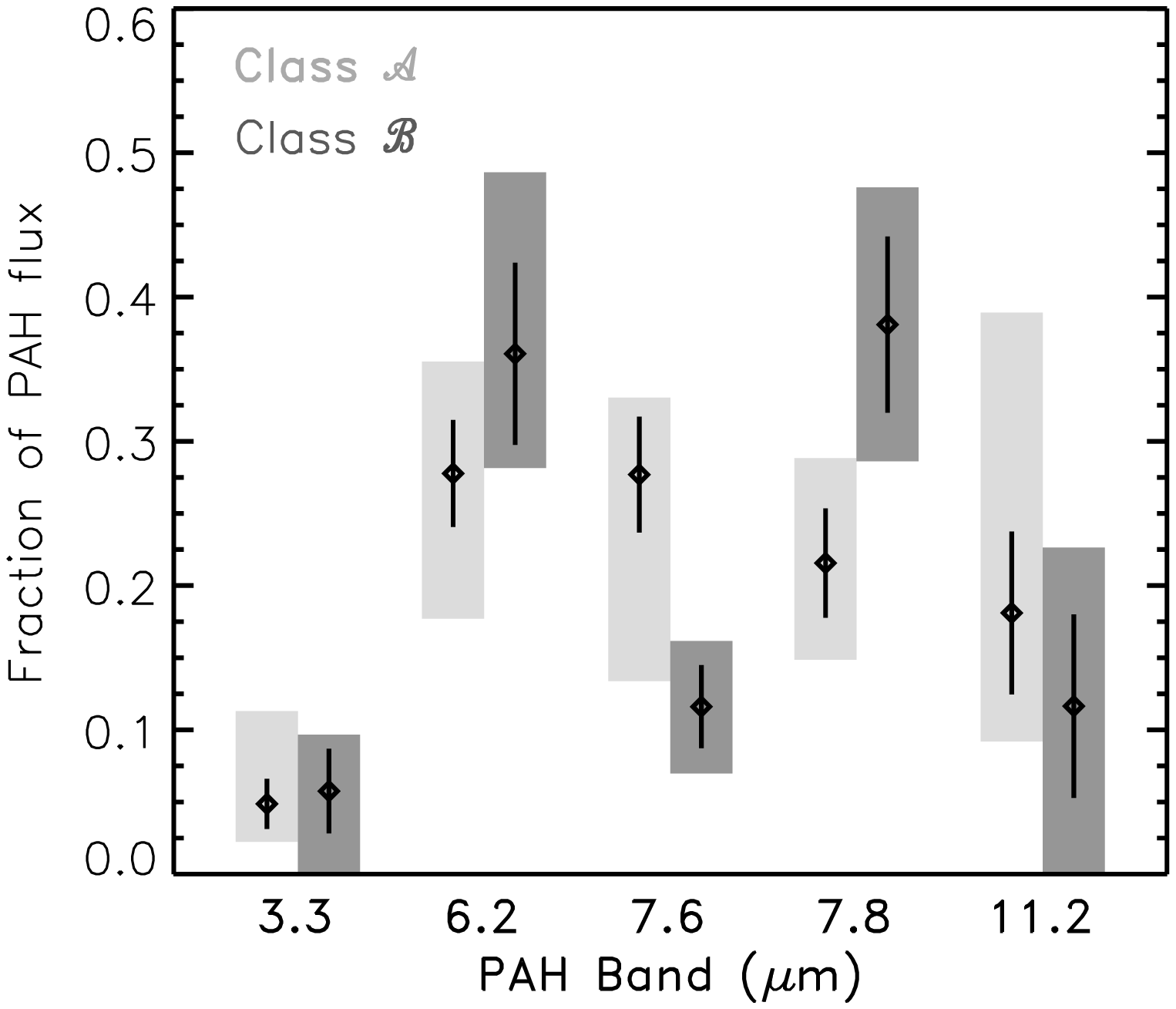,width=240pt,height=240pt}
   \end{picture}
  \end{center}
\caption{An overview of the fraction of the PAH flux emitted in each feature
  for class \ca \, and class \cb. The diamonds crossed by solid
  vertical lines represent the average fraction of the total PAH flux
  emitted in each PAH band over a given class and its standard
  deviation. The filled boxes represents the observed ranges of the
  fraction of PAH flux emitted in each PAH band within a class.
  Light grey represents class \ca \, and dark grey class \cb. For the
  total PAH flux, we considered the 3.3, 6.2, 7.7 and 11.2 \mum\,
  features.}
\label{prof6_meanclass}
\end{figure}

\begin{figure*}[!bht]
  \begin{center}
   \begin{picture}(480,240)(20,0)
   \psfig{figure=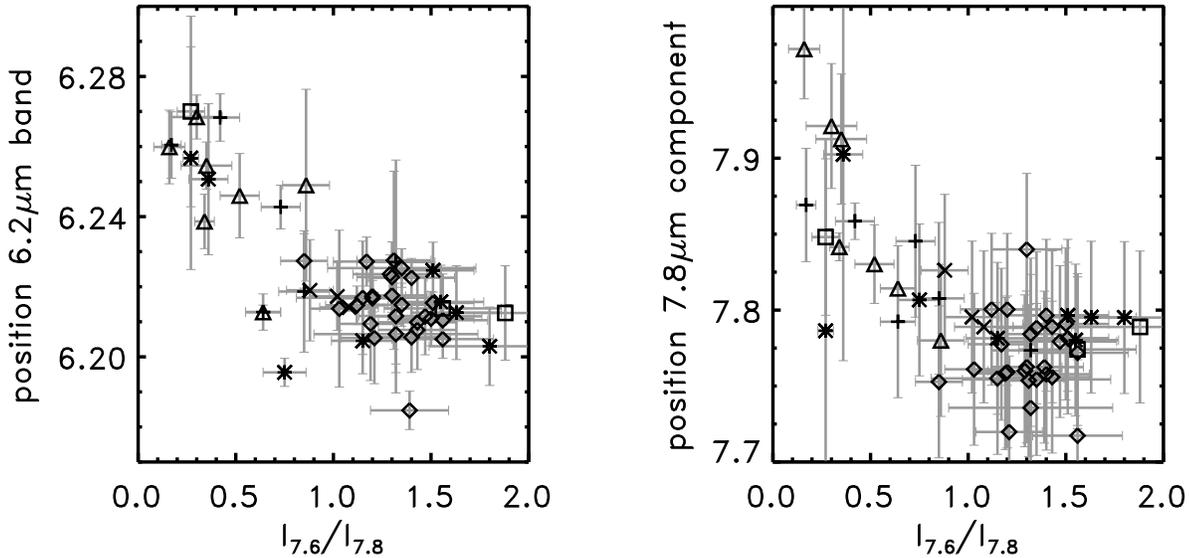,width=480pt,height=240pt}
   \end{picture}
  \end{center}
\caption{The relation between the intensity ratio I$_{7.6}$/I$_{7.8}$
  and the peak position of the 6.2 \mum\, band and the 7.8 \mum\,
  component respectively. Extinction is unimportant for this intensity
  ratio. The symbols are : HII regions +, Herbig AeBe stars
  $\diamond$, RNe/Be/sym $\Box$, Post-AGB stars $\triangle$, PNe
  $\times$ and galaxies $\ast$.}
\label{prof6_posvsratio}
\end{figure*}

For each class, the {\it average} fraction of the total PAH flux,
emitted in each feature, its standard deviation and the total
  range present in our sample are shown in Fig.
\ref{prof6_meanclass}. For the total PAH flux, we only considered the
3.3, 6.2, 7.7 and 11.2 \mum\, features. The fraction of the total PAH
flux emitted in the 7.6 and the so-called 7.8 \mum\, features clearly
depends on the class of the source (see Fig.
\ref{prof6_ratiosvssompah}, left panel and Fig.
\ref{prof6_meanclass}). In contrast, the fraction of the total PAH
flux emitted in the 7.7 \mum\, complex does not differ significantly
for the two classes. Furthermore, the fraction of the total PAH
  flux emitted in the 6.2 and 11.2 \mum\, feature differ for the
  classes \ca \, and \cb \, while a similar fraction is emitted in the
  3.3 \mum\, feature for both classes (see Fig.
  \ref{prof6_ratiosvssompah}, right panel and Fig.
  \ref{prof6_meanclass}). Therefore, it seems that the increase in
I$_{6.2}$/I$_{\Sigma PAH}$ for class \cb \, compared to class \ca \,
is opposite to the behaviour of the 11.2 \mum\, feature. However,
there is no indication of this when plotting I$_{6.2}$/I$_{\Sigma
  PAH}$ versus I$_{11.2}$/I$_{\Sigma PAH}$. In addition, the
  fraction of flux emitted in the 3.3 or 11.2 \mum\, feature does not
  correlate with the fraction emitted in the 6.2 \mum\, feature. So,
this difference in I$_{6.2}$/I$_{\Sigma PAH}$ for the two classes
cannot be directly linked to {\it only one} band (the 3.3, 7.7
or 11.2 \mum\, feature).  Furthermore, sources with an increased
emission in the 6.2 \mum\, feature (class \cb) also have an increased
emission in the 7.8 \mum\, feature and hence a decreased emission in
the 7.6 \mum\, component compared to class \ca\, (see Fig.
\ref{prof6_ratiosvssompah}, right panel).

In Fig. \ref{prof6_posvsratio}, the intensity ratio of the 7.6 to 7.8
\mum\, subfeatures, i.e. I$_{7.6}$/I$_{7.8}$, is plotted versus the
peak position of the 6.2 \mum\, feature and of the 7.8 \mum\,
component respectively. The peak position of the 7.8 \mum\, component
is determined from the position of this subpeak in the 7.7 \mum\,
complex itself and not from the applied Gaussian fits. The error in
this position is given in Table \ref{sample} for entrees in which the
7.7 \mum\, complex peaks longwards of 7.7 \mum. For sources with a
dominant 7.6 \mum\, component,  we estimate the error to be less
  then 0.05 \mum. It is clear that the sources reflect a gradual
variation. In contrast, no clear correlation between the
  I$_{7.6}$/I$_{7.8}$ intensity ratio and the FWHM of the 6.2 \mum\,
  band is present.  

We also checked for correlations between band strength ratios and
the local radiation field G$_0$, however we do not detect any
correlations. Furthermore, the local radiation field of the sample
sources do not show differences between the two classes.  In
addition, we checked for correlations between band strength ratios
and the integrated band-to-continuum ratio, $\Sigma$PAH/cont, in the
6--9 \mum\, region.  This ratio traces excitation conditions and/or
abundance variations. The error on this ratio is dominated by the
continuum determination and is less then 10\%. We took the continuum
described in Sect.  \ref{cont}; hence, the plateau contributes to
the continuum. This ratio is given in Table \ref{sample}. Here also,
we do not detect any correlation. But, sources with the intensity
ratio I$_{7.6}$/I$_{7.8} <$ 1 do not have a high $\Sigma$PAH/cont
ratio ($<$ 0.45) whereas sources where I$_{7.6}$/I$_{7.8} >$ 1 span
a range in $\Sigma$PAH/cont from 0 to 1.2.

\section{The Infrared Emission Features and PAHs}
\label{spec}

\begin{figure}[!tbh]
  \begin{center}
   \begin{picture}(240,240)(10,10)
   \psfig{figure=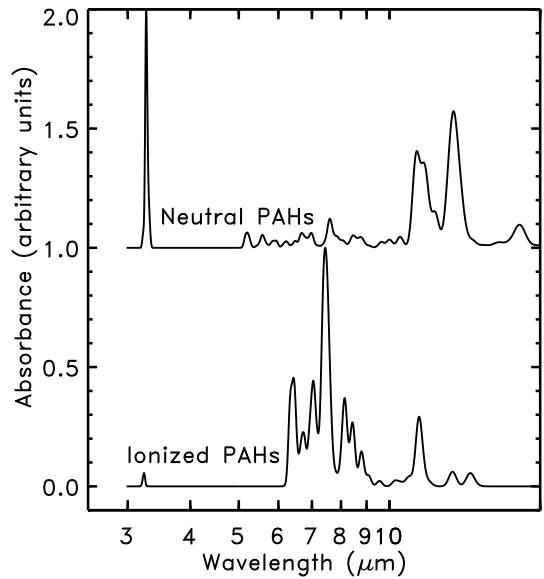,width=240pt,height=240pt}
   \end{picture}
  \end{center}
\caption{The absorption spectrum of a mixture of neutral PAHs (a)
compared to the spectrum of the same PAHs in their positive state (b).
This comparison shows that, for PAH spectra, ionisation has a much
greater influence on relative intensities than on peak frequencies,
with the features in the 6 to 10 \mum \, region substantially enhanced with
respect to the rest of the spectrum \citep[Figure adapted from][]{Allamandola:modelobs:99}.}
\label{neutralvsion}
\end{figure}

The IR emission features at 3.3, 6.2, 7.7, 8.6, and 11.3 \mum \, are
now generally thought to arise from vibrationally excited PAHs due to
their similarity with the spectra of PAHs taken under conditions which
match the salient characteristics of the interstellar environment.
One of the early pivotal results of the laboratory and theoretical
studies on PAHs reported over the last decade is the remarkable effect
ionisation has on the infrared spectra \citep[][and
ref. therein]{Szczepanski:labcations;93, Langhoff:neutionanion:96,
Kim:gasphasepyrenecation:01, Hudgins:tracesionezedpahs:99}.  While PAH
characteristic frequencies are only modestly affected by ionisation,
the influence on intensity is striking - particularly in the 5 to 10
\mum \, region (Fig. \ref{neutralvsion}).  The bands grow from the
smallest features in neutral PAH spectra to become the dominant bands
in ionised PAH spectra.  The CC stretching vibrations grow in
intensity because, upon ionisation, the charge distribution changes
significantly with the CC skeletal vibration, creating a strong
oscillating dipole whereas the oscillating dipole for the CH motions
and their corresponding bands are reduced.  
Fig. \ref{neutralvsion} illustrates this effect, which has been found
for all PAHs measured in the laboratory to date. However, this is
mostly limited to PAHs with less than 48 C-atoms. Larger PAHs are only
now being studied in the laboratory. Preliminary laboratory results
for PAHs containing up to 60 C-atoms also show this behaviour upon
ionisation. Theoretical studies of large species support this
enhancement to much larger sized PAHs. For example,
\citet{Bauschlicher:C96:02} has shown that this holds for
C$_{96}$H$_{24}$.  Astronomically, the observation that the 6.2 and
7.7 \mum \, features, the focus of this paper, are the most intense of
the interstellar emission band class, is a clear indication that
ionisation is an intrinsic characteristic of interstellar PAHs. 
The 3.3 \mum\, feature on the other hand is characteristic for neutral
PAHs. Hence, both ionised and neutral PAHs contribute to the
interstellar IR emission spectrum.

\begin{figure}[!tbh]
  \begin{center}
   \begin{picture}(240,240)(10,10)
   \psfig{figure=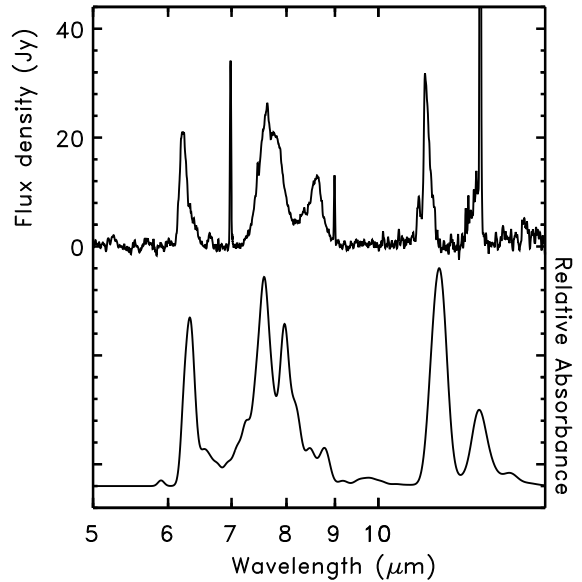,width=240pt,height=240pt}
   \end{picture}
  \end{center}
\caption{Comparison of a typical ISM infrared emission spectrum (a)
  with a composite absorption spectrum generated by 
co-adding the individual spectra of 11 PAHs (b). The interstellar
spectrum is that of IRAS~23133.  The individual
spectra were calculated using experimentally measured frequencies and
intensities and assigning a 30 cm$^{-1}$ FWHM Gaussian band profile,
consistent with that expected from the interstellar emitters \citep[e.g.][]{Hudgins:spacing:99}.}
\label{orionfit}
\end{figure}

There have been many comparisons between PAH spectra and the
interstellar spectra over the years.  As the observational tools have
become more sensitive and the laboratory techniques more appropriate
to the interstellar case, the fits have become more revealing about
the different PAH populations in different regions and this, in turn,
has yielded further insight into conditions in the emitting regions.
Fig. \ref{orionfit} shows a fit to the emission from IRAS~23133 measured
with ISO by spectra now available in the Ames PAH IR spectral database
(\texttt{http://web99.arc.nasa.gov/\~{ }astrochem/pahdata/}\linebreak
\texttt{index.html}).  Although
the fit is striking, thanks to the quality of the new ISO spectra,
important differences become apparent which shed
further light on the interstellar PAH population.  The following
differences can be seen in going from shorter to longer wavelengths.
The laboratory band near 6.2 \mum \, falls slightly to the red of the
interstellar feature, laboratory components centred near 7.7 \mum \, do
not precisely match the interstellar profile at this position, the 8.6
\mum \, laboratory band is weaker with respect to the other features in the
spectrum than is the case for IRAS~23133, and the laboratory component near
11.2 \mum \, lies to the red of the interstellar feature.  The differences
in the 8.6 and the 11.2 \mum \, regions are discussed elsewhere
\citep[][Janssen, Janssens and Swansen 2002, in prep.]{Hony:oops:01}.  Here
we focus on the interstellar features near 6.2 and 7.7 \mum.

\subsection*{Feature assignments in the 5 to 10 \mum \, region}

The 5 to 10 \mum\, region encompasses frequencies which originate from a
variety of PAH molecular vibrations.  Pure CC stretching motions
generally fall between about 6.1 \mum \, and 6.5 \mum, vibrations involving
combinations of CC stretching and CH in-plane bending modes lie
slightly longward, between roughly 6.5 \mum \, and 8.5 \mum, and CH in-plane
wagging vibrations give rise to bands in the 8.3 \mum \, to 8.9 \mum \, range.
While the well-known interstellar features at 6.2, 7.7, and 8.6 \mum \,
dominate this range, there are at least four weak interstellar bands
in this region as well, centred near 5.2, 5.6, 6.0 and 6.8 \mum.
Their correlation with the major features indicates that they too
originate from the interstellar PAH family.  The 5.2 and 5.6 \mum \,
features most likely correspond to combinations and overtones
involving the CH out-of-plane fundamental vibrations which fall
between 11 and 13 \mum; the 6.0 \mum \, feature likely indicates a carbonyl
($>$C=O) stretch of an oxygenated PAH (a quinone); and the 6.8 \mum \, band
probably corresponds to a weak aromatic CC stretching - CH in-plane
bending combination mode of PAHs as for example in the fluoranthenes, or the
aliphatic -CH$_2$- or -CH$_3$ deformation in a methyl or ethyl side-group
attached to a PAH.\\

 There are also other plausible interstellar PAH-related species
that are likely to be important in the emission zones and which should
be considered. Some examples include PAH clusters and PAH complexes
with metals such as iron (metallocenes). Furthermore, one can question
whether PAHs remains planar as they grow. And, if not, how does the
3-dimensional shape influence the IR spectrum? It is noteworthy in
this regard that simple fullerenes do not have IR characteristics
which coincide with any prominent structure in the interstellar
emission spectra discussed here \citep{Moutou:c60:99}. Further, and in spite
of much effort, there has not been any report of IR active transitions
in carbon nanotubes, very large curled aromatic networks. If such
curled aromatic networks (very, very large PAHs) do possess infrared
activity, it appears to be very weak. Thus here we focus on the
PAH molecules measured in the laboratory or theoretically
calculated. 
   
The shifts in the profiles shown by the observations presented above
(see Table \ref{sample}) are much larger than the binsize
corresponding to the obtained resolution of the data. Hence, they are
chemical in nature, arise from local excitation conditions and are not
due to doppler broadening and shifting. Therefore, these shifts and
profile variations provide important new insight into the variations
of the interstellar PAH populations in the different environments.
Interpretation of these new spectral aspects require probing deeper
into the details of PAH spectroscopic properties in this region than
heretofore. Here we consider this new information as it affects first
the 6.2 \mum \, region and then the 7.7 \mum \, region.

\subsection{The position of the PAH CC stretching band near 6.2 \mum
  \,: experiment and theory}
\label{lab62}

There are several properties which determine the precise peak positions
of the infrared active bands which correspond to pure CC stretching
vibrations in any given PAH.  These include molecular size, molecular
symmetry, and molecular heterogeneity.  The roles each of these
play in determining the position are discussed below. 
 The charge of the PAH
molecules also shifts the band position; but this trend is not
systematic \citep[][Bakes~E., private
communication]{Bakes:modelI:01}. In addition, dehydrogenation
influences the band position \citep{Pauzat:dehydro:97}. However, based
upon the CH out-of-plane bending modes, \citet{Hony:oops:01} conclude that
dehydrogenation has little influence on the observed interstellar PAH
spectrum.

\paragraph{Molecular size}

\begin{figure}[!htb]
  \begin{center}
   \begin{picture}(240,240)(0,10)
   \psfig{figure=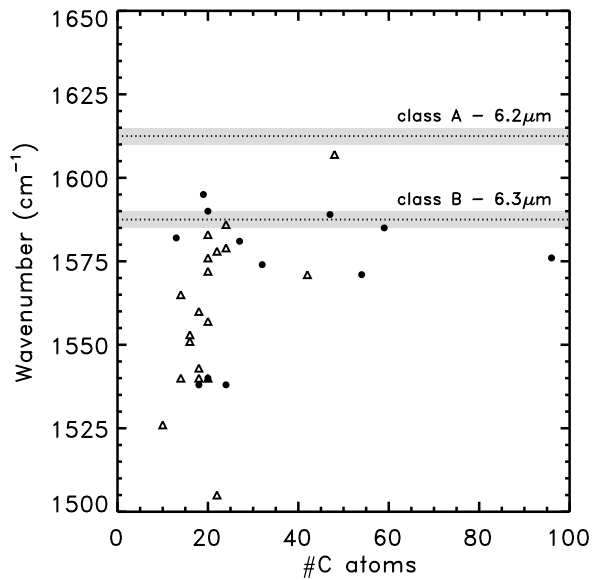,width=240pt,height=240pt}
   \end{picture}
  \end{center}
\caption{A plot of the dominant feature position in the PAH
CC stretching region as a function of molecular size.  The positions
and typical FWHM of the interstellar 6.2 and 6.3 \mum \, components
discussed in Sect. 
\ref{f62} are indicated by the horizontal shaded bands.  Open symbols
correspond to experimentally measured positions.  Solid symbols are frequencies
computed using B3LYP quantum calculations \citep[][Bauschlicher,
Hudgins, and Allamandola 2002 (in
prep.)]{Hudgins:closedshellpahcation:01, Bauschlicher:C96:02}. 
Theoretical positions typically have uncertainties of 5-10
wavenumbers, with a few bands showing shifts of 15 to 20 wavenumbers. }  
\label{catomsvspos62}
\end{figure}

At the small end of the PAH size distribution (C$_{10}$ to
$\sim$C$_{30}$), the peak wavelength of the dominant CC stretching feature
decreases steadily with molecular size (Hudgins and Allamandola,
1999).  The data shown in Fig. \ref{catomsvspos62}, which has been updated to
include more PAHs than available in the original study, shows
this trend.  
  
However, Fig. \ref{catomsvspos62} shows that the trend does not
continue ad-infinitum but seems to die out for molecules with more
than about 30 to 40 carbon atoms, falling slightly longwards of 6.3
\mum.  This behaviour is consistent with the limit of
graphite which shows an emission mode at 6.3 \mum \,
\citep{Draine:graphitegrains:84}. Thus, as the size grows, the
influence of boundary conditions on these pure CC stretching
vibrations within the PAH carbon skeleton vanishes when the peak
wavelength of this mode in graphite is reached.
Indeed, it is a general chemical rule that within any molecule, the
further a given bond or chemical subgroup is from the site of a
modification, the smaller the effect exerted by that change on the
properties of the specific bond or subgroup.  This goes for a
molecule's fundamental vibrational frequencies as well as its chemical
properties. For the smallest PAHs, addition of
even a single ring constitutes a significant modification to the
carbon skeleton and strongly influences its pure CC stretching
vibrational frequencies. However, as PAH size increases, an
ever-increasing fraction of the molecule's carbon skeleton lies "far" from
the site of any particular modification and is thus increasingly
insensitive to that modification. As a result, the corresponding
fundamental CC stretching modes that arise within that skeleton are
progressively less perturbed resulting in a levelling off of their
frequencies above a certain size ($\sim$30 C atoms as estimated from
Fig. \ref{catomsvspos62}).

Importantly from the astrophysical perspective, Fig.
\ref{catomsvspos62} shows that the maximum wavenumber falls some 20
cm$^{-1}$ short of 6.2 \mum, the peak 
position of the Type A interstellar bands discussed in Sect.
\ref{f62}.  This mismatch between PAH band position and the
interstellar feature is only worsened by the approximately 10 cm$^{-1}$ red
shift which occurs for emission from a vibrationally excited PAH
\citep[][]{Cherchneff:pah+uvlaser:89,Flickinger:T:90,Brenner:benz+naph:92,Colangeli:T:92,Joblin:T:95,Cook:excitedpahs:98}.
Thus, while the correlation between molecular size with 
band position is confirmed by new experimental and theoretical data,
for the highly symmetric pure PAHs considered here, the strongest,
infrared active, pure CC stretching mode cannot reproduce the position
of the 6.2 \mum \, interstellar feature and other factors which can
slightly shift this frequency must be considered. 

\paragraph{Molecular symmetry}

Another property which can influence IR activity
is molecular symmetry.  In large, symmetric molecules there are many
vibrations which correspond to very weak or infrared forbidden
transitions.  These vibrations are weak or IR inactive in highly
symmetric species because the oscillating dipoles arising from atomic
motions at one position in the molecule are cancelled by identical,
but oppositely oriented dipoles arising from identical, oppositely
phased motions occurring elsewhere in the molecule.  Reducing
molecular symmetry eliminates some of this oscillating dipole
cancellation and once IR forbidden modes can become IR active.  This
behaviour is expected independent of the means by which symmetry is
broken.  For large, highly symmetric molecules such as circumcoronene
(C$_{54}$H$_{18}^+$, symmetry = D$_{6h}$) and circum-circumcoronene
(C$_{96}$H$_{24}^+$, symmetry = D$_{6h}$) included in Fig.
\ref{catomsvspos62}, the calculations indicate 
that there are several infrared inactive CC stretching
vibrations that fall between about 6.15 and 6.25 \mum, (1626 and 1600
cm$^{-1}$) positions which
overlap the interstellar position.  Since there is no reason to expect
that interstellar PAHs would be so highly symmetric, we have explored the
influence of symmetry breaking on the IR spectrum to assess the
possibility that asymmetric PAHs would show strong IR activity at
shorter wavelengths, closing the gap between the PAH spectra and the
6.2 \mum \, interstellar feature.  Several different cases were tested and
all show only marginal effects (Bauschlicher, Hudgins, and
Allamandola 2002, in prep.).  For example, by reducing the symmetry of
the circumcoronene cation by removing first one and then two of the
peripheral rings, the peak position of the strongest pure CC vibration
does not shift much if at all.  It falls at 6.36 \mum \, (1572
cm$^{-1}$) in both circumcoronene and circumcircumcoronene with one ring
removed and only slightly shifts to 6.33 \mum \, (1580 cm$^{-1}$) in
circumcoronene with two rings removed. Thus, we conclude that symmetry
breaking alone cannot account for the difference between the
interstellar feature and the PAH spectra considered to now.

\paragraph{Molecular heterogeneity}

\begin{figure}[!tbh]
  \psfig{figure=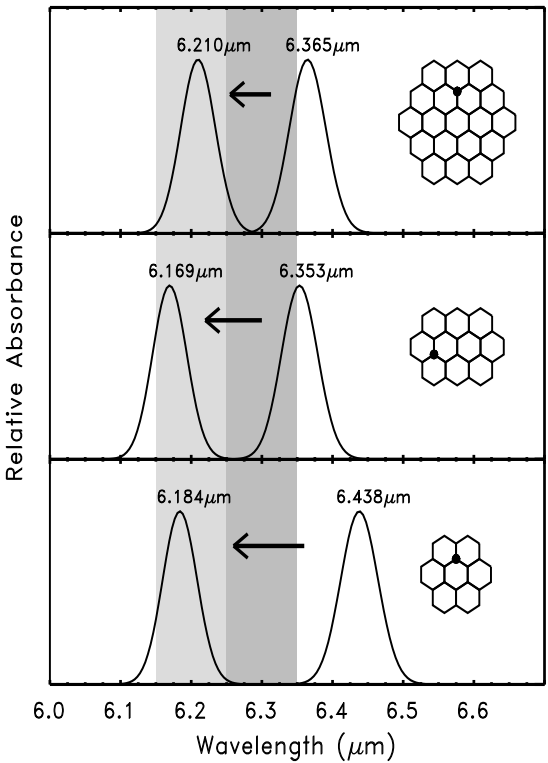,width=220pt,height=320pt}
\caption{The effect of PAH nitrogenation on the positions of the
  strongest band in the CC stretching region for the three PAHs
  coronene, ovalene, and circumcoronene. The right profile indicates
  the position for the unsubstituted PAH, the left for the nitrogenated
  species.  The position of nitrogen substitution is indicated by the
  filled circle
  in the structures.  The vertical grey shaded bands indicate the
  position and width of the interstellar type A and B bands discussed
  in Sect. \ref{f62}.}
\label{nitrogenation}
\end{figure}

Another means by which IR activity can be induced in otherwise weak or
forbidden transitions is by the introduction of a hetero atom into the
symmetric molecular structure.  This alters the molecule in two ways:
it lowers the symmetry and modifies the electronic distribution.  To
investigate this influence on PAH spectroscopy, Bauschlicher, Hudgins,
and Allamandola (2002, in prep.) have studied the effects of
introducing different hetero-atoms into different positions within the
carbon network of several PAHs.  This affects, in particular, the
position of the 6.2 \mum\, band.  A few examples are presented in Fig.
\ref{nitrogenation}.  Here the positions of the strongest band in this
region are shown for the cations of the pure PAHs coronene, ovalene,
and circumcoronene and their counterparts which have one of the carbon
atoms replaced by a nitrogen.  In all cases nitrogen substitution
shifts the strongest CC stretching vibration to the 6.2 \mum \,
interstellar band position.  Further within a given PAH, the shift
depends somewhat on the "depth" of the substitution in the carbon
skeleton.  For example, when nitrogen is substituted for a carbon atom
on the outermost ring of the circumcoronene cation, the strongest band
in the CC stretching region falls at 6.16 \mum; when this substitution
is made on the next innermost ring (shown in Fig.
\ref{nitrogenation}), it falls at 6.21 \mum; and when on the innermost
ring it falls at 6.19 \mum.  Bandshift behaviour when more than
  one nitrogen is substituted is still to be determined. For singly
  substituted PAHs, this peak shift has two causes.  First, the
insertion of nitrogen decreases molecular symmetry, increasing IR
activity in modes which are normally weak because of the higher
symmetry as described above.  Second, the strongly electro negative
nitrogen atom modifies the fixed charge distribution on the carbon
network and induces stronger IR activity for these vibrations since
the magnitude of the oscillating dipole increases.  This behaviour can
be expected independent of the means by which the charge distribution
is fixed.

Bauschlicher, Hudgins, and Allamandola (2002, in prep.) have also
considered oxygen and silicon atom substitution.  Nevertheless, while
these too show a similar induction of IR activity close to 6.2 \mum,
when taking chemical considerations into account, nitrogen
substitution is still the most attractive candidate. First, nitrogen
can be incorporated anywhere within the ring structure without
compromising the aromatic stability of the $\pi$ electron
network. Second, oxygen substitution does not have this
advantage. Oxygen or O$^+$ cannot form four bonds and so
oxygen is not found in the middle rings. Hence, oxygen would be most
stable only at the edge positions, positions which would be more
reactive and subject to chemical attack. Third, although silicon can form four
bonds, the C-Si bond length is larger then the C-C bond length and
hence the aromatic rings will be disturbed by silicon
substitution. Moreover, silicon has a much lower cosmic abundance then
does nitrogen. 
Thus, at this stage in our understanding, large PAHs
containing some nitrogen seem most plausible to account for the 6.2
\mum \, band position. As mentioned before, there are other plausible interstellar
PAH-related species in which one might induce activity at this
position and these must be investigated before a firm conclusion can
be drawn.  Some examples include large PAHs with
uneven and irregular structure, PAH clusters and PAH complexes with
metals such as iron (metallocenes).

\subsection{The position of the PAH CC stretching/CH in-plane bending
  feature near 7.7  \mum \,: experiment and theory} 
\label{lab77}

\begin{figure}[!h]
  \begin{center}
   \begin{picture}(240,240)(0,10)
   \psfig{figure=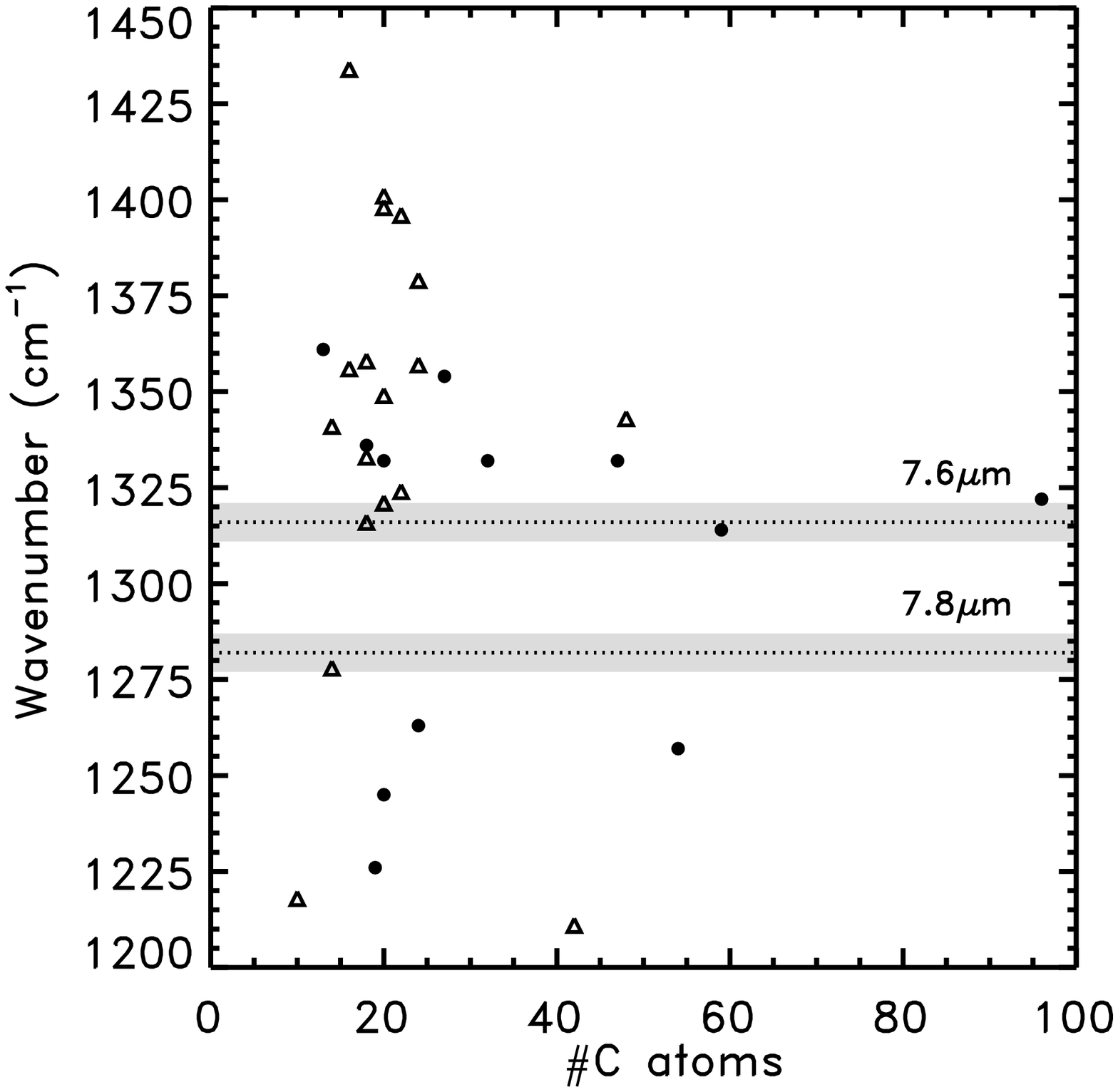,width=240pt,height=240pt}
   \end{picture}
  \end{center}
\caption{A plot of the dominant feature position in the 
    wavelength region of the 7.7 \mum\, complex as a function of
  molecular size.  The positions and typical FWHM of the interstellar
  7.6 and 7.8 \mum \, components of class \ca\, discussed in Sect.
  \ref{f77} are indicated by the horizontal shaded bands. The two
  components of the 7.7 \mum\, complex in Class \cb\, peak at 7.6 and
  from 7.8 to 8.0 \mum\, respectively.  Open
  symbols correspond to experimentally measured positions.  Solid
  symbols are frequencies computed using B3LYP quantum calculations
  \citep[][Bauschlicher, Hudgins, and Allamandola 2002 (in
  prep.)]{Hudgins:closedshellpahcation:01, Bauschlicher:C96:02}. 
    Theoretical positions typically have uncertainties of 5-10
    wavenumbers, with a few bands showing shifts of 15 to 20
    wavenumbers. }
\label{catomsvspos77}
\end{figure}

The CC stretching/CH in-plane bending vibrations of most singly
ionised PAHs possess at least one very strong feature between about 7.2
and 8.2 \mum.  Fig. \ref{catomsvspos77} shows a graph of the
frequencies for the strongest of these modes plotted versus the carbon
number for the same molecules considered in Fig.
\ref{catomsvspos62}.  In this case (Fig. \ref{catomsvspos77}) there is 
no clear relationship between frequency and size in contrast to the
behaviour for the pure CC stretch (Fig. \ref{catomsvspos62}).  While a
few of the smallest molecules do have vibrations which fall at the lowest
frequencies measured for these modes - suggesting such a correlation -
similarly sized molecules possess some of the highest vibrational
frequencies. Furthermore, one of the largest PAHs in the laboratory database,
hexabenzocoronene (C$_{42}$H$_{24}$), has the lowest frequency determined to
date.  Here peak positions for small and large PAHs are intermingled. 

In addition, Fig. \ref{catomsvspos77} reveals that there is also a gap
between about 7.6 and 7.8 \mum \, (1316 and 1282 cm$^{-1}$), with a
clustering of the data between 1320 and 1355 cm$^{-1}$ (7.4 and 7.6
\mum).  The data in Fig.  \ref{catomsvspos77} suggests that the mean
PAH CC stretching/CH in-plane bending frequency lies near 1335
cm$^{-1}$ (7.5 \mum).  This behaviour is important in view of the
results presented in Sect. \ref{f77} which show that two components
dominate the interstellar emission in this region, one peaking near
7.6 \mum \, (class A$'$) , the other longwards of 7.8 \mum \, (class
B$'$).  When one takes the roughly 10 cm$^{-1}$ red shift into account
for emission, it is apparent that this family of PAHs can readily
reproduce the 7.6 \mum\, component, but not the dominant 7.8 \mum\, component.

As discussed above for the pure CC stretching features near 6.25 \mum,
we have theoretically considered how molecular symmetry and
heterogeneity influence the peak position of the 7.7 \mum\,
  complex in order to understand what molecular properties might be
responsible for the interstellar position.  Breaking the high symmetry
of PAH molecules, either by removing rings or by substituting a
nitrogen atom at several different positions within the carbon network
again produces IR activity in modes which were weak or completely
forbidden (Bauschlicher, Hudgins, and Allamandola 2002, in prep.).  In
this case however, while a few examples are found in which intense
bands fall in this gap, these results seem random and no clear
structural relationship has yet emerged. Thus, this sort of molecular
modification does not account for the interstellar 7.8 \mum \,
component.

 Furthermore, as with the 6.2 \mum\, feature,
  dehydrogenation and the charge of the PAH 
molecules do not play a major role in determining the peak position
(see Sect. \ref{lab62}).

In summary, the 7.6 \mum \, (class A$'$) interstellar band is consistent
with a mixture of large and small, pure and hetero-atomic PAHs.  These
PAHs also contribute emission at 7.8 \mum \, as illustrated in Fig.
\ref{orionfit}. However, the PAHs so far considered reveal a dearth of
strong spectral features longwards of 7.7 \mum \, and thus have
difficulty reproducing the class B$'$ 7.7 \mum\, band. There are only a few 
PAHs in the database which have their strongest band at 7.8 \mum \, and the
data is insufficient to make any generalisations concerning the
properties which cause strong IR activity at this position.  As above,
there are several other PAH related species that are likely to be
important in the emission zones and which should be considered.

\section{Astronomical implications}
\label{discussion}

From the wealth of IR spectra, it is clear that the
UIR bands at 3.3, 6.2, "7.7", 8.6 and 11.2 \mum\, represent a single
class of spectral features that come and go together. 
The ISO observations presented here in Sect. \ref{f62}, \ref{f77} and
\ref{f86}, show that the 
peak positions and profiles of the 6.2, "7.7" and 8.6 \mum\, features vary
significantly from source to source. Moreover, as discussed in \S
\,\ref{classes}, these variations in the different bands are
correlated with each other and with the type of object considered
(i.e. HII region, YSO, Post-AGB star and so on).  These variations
stand in marked contrast to the behaviour of 3.3
and 11.2 \mum \, bands.  While the latter modes show some variation in
profile, their peak position is relatively invariant with both 
bands wandering by only 0.2\% 
\citep[][Peeters et~al. 2002, in prep.; van Diedenhoven
et~al. 2002, in prep.]{Tokunaga:33prof:91,Hony:oops:01}.
Furthermore, while the 3.3 and 11.2 \mum \, band intensities are
correlated with each other \citep{Hony:oops:01}, their strengths do
not correlate well with that of the 6.2 and 7.7 \mum \, features.

\subsection{The observed trends}

It is shown that class \ca\, and \cb\, have distinctly
different profiles, peak position and relative intensities of the
features in the 6--9\mum\, region. We attribute these differences
between the two classes to variation in the PAH families present
around the sources; i.e. different molecular mixtures are found
around the sources of the two classes whose combined IR spectral
fingerprints change the overall PAH spectrum in the 6--9 \mum\,
region significantly. 

Class \cb\, corresponds to regions where G$_{0} > $ 10$^4$. For
such high G$_{0}$, very small grains (VSG) emit in the 6-9 \mum\,
region \citep{Cesarsky:sileminorion:00}. Since the range in G$_{0}$
present in class \cb\, is also present in class \ca, these VSG are
not responsible for the observed differences.  Furthermore, these
differences in the observed emission features and hence in the PAH
families is not directly related with G$_{0}$. 

\citet{Hony:oops:01} derived the molecular structure of the PAHs based
upon the 10-15 \mum\, spectra and concluded that the PNe contain
large, compact PAHs with long straight edges while HII regions contain
smaller or more irregular PAHs. Clearly, the HII regions and
non-isolated Herbig Ae Be stars all belong to class \ca\, and have a
weak I$_{11.2}$/I$_{12.7}$ ratio. But, for the PNe considered in
\citet{Hony:oops:01}, no link is found between the
I$_{11.2}$/I$_{12.7}$ intensity ratio and the profile classes. This
supports the above mentioned observations that the CH modes seems to
behave independently of the CC modes.

\subsection{The feature assignments and carriers}

In general, the peak position of the PAH vibrations may be affected by
charge (anion, cation, and neutral states), size, symmetry and molecular
structure and heterogeneity.  It is therefore important to understand
the precise transitions involved in producing these interstellar
features if we are to fully interpret what these transitions tell us
about the nature of the carriers, their history and the local physical
environment.     

\subsubsection{The 6.2 \mum\, feature}
\label{feat62}

The observations presented here (Figs. \ref{varA} through
\ref{gauss}) show that the peak 
position of the interstellar PAH CC stretching band varies between 6.2
and 6.3 \mum.  The interstellar feature seems more red-shaded
  (i.e. asymmetric with a red wing)
when it peaks at 6.2 \mum \, (Class A emitters, Table 2) then when it
peaks near 6.25 \mum \, (Class B emitters, Table 2).  For Class A
emitters, the observed half-widths (0.11 \mum, 28 cm$^{-1}$) are close
to the expected intrinsic line width of an emitting PAH molecule or
cluster (roughly 30 cm$^{-1}$), for Class B emitters the observed
widths (0.15 \mum, 38 cm$^{-1}$) are slightly larger and Class C
emitters have the smallest widths (0.099 \mum, 25 cm$^{-1}$).  

The variation in the observed peak positions and profiles are
interwoven and may reflect the effects of a class of PAHs, each with
its individual peak position, or the effects of anharmonicity inherent
to the emission process in higher vibrationally excited PAHs, or
both. 

\subsubsection*{Emission by a PAH family} 

As discussed in Sect. \ref{lab62} and illustrated in
Fig. \ref{catomsvspos62}, pure PAHs (i.e. those containing only carbon
and hydrogen) composed of $>$20-30 C atoms can reproduce the position
of the interstellar 6.3 \mum\, component but cannot account for the
6.2 \mum\, component. Based on the currently-available experimental
and theoretical data, the most probable carriers of the 6.2 \mum\,
component are hetero-atom (e.g. N, O, Si) substituted PAHs, PAH
clusters and/or PAH complexes with a metal atom such as iron
(i.e. metallocenes). Interactions in such species may alter a PAH
molecule in two ways : they may lower the molecule's symmetry and they
may alter the electron distribution within the molecule. Both of these
effects may shift the dominant CC stretching modes toward shorter
wavelengths. 

In summary, the observed profiles of
the 6.2 \mum\, band complex can be a "blend" of emission bands by
different carriers with slightly shifted peak position for the main CC
mode. Pure PAHs represent the 6.3 \mum \, component and
substituted PAHs or PAH-like species emits the 6.2 \mum\, component.
The profile of the 6.2 \mum\, band complex reflects then the relative
abundance of the different species in the PAH population and so different
sources contain different amounts of each type of 
species. Typically, the PAH population in Post-AGB objects, PNe and
isolated Herbig AeBe stars is skewed toward pure-C PAHs, while HII
regions, RNe, non-isolated Herbig AeBe stars and galaxies have a
dominant contribution from substituted complexed PAHs. However, the PAH family
in some PNe also contains an important substituted PAH population. NGC~7027 is a
case in point. This interpretation raises interesting questions on the 
origin and evolution of circumstellar and interstellar PAHs which will
be addressed in Sect. \ref{formevol}. In principle, the family could
consist of two (distinct) species only. However, we consider that unlikely.

\subsubsection*{Anharmonicity} 
\label{anharm}

It is also possible to interpret the observed (red
shaded) profiles in terms of emission from a single PAH (or a
collection of PAHs with very similar peak position) which is (are) highly
vibrationally excited.  If a molecule is 
sufficiently highly vibrationally excited, emission from levels above
the first excited state become important.  Due to the anharmonic
nature of the potential well, these band spacings become smaller and
smaller the higher up the vibrational ladder one samples and emission
between these levels falls slightly and progressively to the red
producing a long wavelength wing reminiscent of the observed wing
\citep{Barker:anharm:87}. In addition, anharmonic coupling of the
emitting mode with other modes also shifts the peak position of the
emitting band to lower energies. Integrating over the energy cascade
as the emitting species cools down will then in a natural way give rise
to a red shaded profile \citep{Barker:anharm:87, Pech:prof:01,
Verstraete:prof:01}.

\citet{Pech:prof:01} and \citet{Verstraete:prof:01} have modelled the
IR emission spectrum 
of PAHs based upon extensive laboratory studies of the shift in peak
position as a function of temperature of the emitter which is a direct
measure of the anharmonicity \citep{Joblin:T:95}.
They obtained excellent fits to the red shaded appearance of the
observed profiles of the 3.3, 6.2 and 11.2 \mum\, bands. However, for
the 6.2 \mum\, band no good fit to the peak position is obtained for
sources with class \ca\, spectra, because the laboratory and theoretical
studies have been limited to pure-C PAHs. Nevertheless, the principle
remains the same and small, highly vibrationally excited,
N-substituted PAHs are also expected to have red shaded emission profiles.

Within this interpretation, the class \ca\, profiles which are highly
asymmetric are due to a relatively limited number of small highly
excited N-substituted complexed PAHs. The class \cc\, profiles, which
peak at 6.3 \mum\, and are fairly symmetric (cf. Fig.  \ref{varC}),
are carried by much less highly excited pure-C PAHs. The "low"
excitation of these pure-C PAHs may reflect either the cool nature of
the illuminating source in these two Post-AGB objects (CRL~2688 and
IRAS~13416) or it may reflect an on-average larger size of the
emitting species, or both. Class B profiles show a less pronounced
blue rise and a less pronounced red wing. We note that class \cb\,
sources typically have a higher I$_{11.2}$/I$_{12.7}$ intensity ratio
than class \ca\, sources \citep{Hony:oops:01}. A high
I$_{11.2}$/I$_{12.7}$ intensity ratio indicates the dominance of
rather large ($\sim$ 150 C-atoms) PAHs \citep{Hony:oops:01} and
anharmonicity effects are expected to be smaller for such large PAHs
\citep{Pech:prof:01}.  The YSO, BD+40 4124, which belongs to class
\ca, shows a Gaussian profile, possibly indicating that the
photochemical survival of the fittest (i.e. most robust) members of
the PAH family has been of great importance in this source (Van
Kerckhoven et~al. 2002, in prep.).\\

The 6.2 \mum\, band of many of these sources have very similar
profiles, irrespective of the harshness of the illuminating radiation
field - as measured by its strength or effective temperature. As
exemplified by the model study of \citet{Verstraete:prof:01}, this
indicates that the typical size of the emitting PAH is larger in
regions which are illuminated by hotter stars. This coupling between
size and the "colour" of the illuminating radiation field may be a
natural consequence of emission from a family of PAH species; that is,
the smallest size of PAHs which can survive in a radiation field will
depend on the average photon energy in the illuminating FUV field. To
phrase it differently, both the minimum size and the profile of the
6.2 \mum\, band may be a measure of the average excitation of the
emitting PAH \citep{Verstraete:prof:01}. The more symmetric profiles
of class \cc\, may then reflect that the PAHs in these cool Post-AGB objects
have not yet been exposed to harsh radiation fields and their
composition still reflects the condition during their formation at high
temperatures. As a corollary, this implies that there are no small
PAHs ($\leq$ 25 C atoms) with peak positions longwards of 6.3
\mum. The coolness of the radiation field and symmetric profiles imply
then that the PAHs in these sources are only moderately excited.

\subsubsection{The 7.7 \mum\, complex}
\label{feat77}

As discussed in the spectroscopy section (Sect. \ref{lab77}), the
observed 7.7 \mum\, profiles of class A$'$ can be remarkably well
reproduced by either pure-C or N substituted PAHs. The dominance of the
7.8 \mum\, band in class B$'$ profiles, however, remains an enigma.

When investigating other possible carriers for the 7.8 \mum \,
component, one should bear in mind the following observational facts.
First, the 7.7 \mum\, complex is always observed together with the
other UIR bands. Second, the position of the 7.8 \mum\, component
correlates with the observed intensity ratio
I$_{7.6}$/I$_{7.8}$. Third, the strength of the 7.8 \mum\, component
is correlated with the strength of the 6.2 \mum\, feature. In
addition, the different classes of the 6.2 and 7.7 \mum\, features are
directly linked with each other. Hence, the carrier of the 7.7
\mum\, complex and in particular of the 7.8 \mum\, component should be
related to the carrier of the other UIR bands.

Other carriers have been proposed to explain the UIR bands. 
Proposed solid state carriers are QCC \citep{Sakata:QCC:84}, soot
\citep{Allamandola:autoexhaust:85}, coal \citep{Papoular:coalmodel:89}, HAC
\citep{Colangeli:amorphc:95, Scott:HAC:97} and nano-sized carbon grains
\citep{Herlin:nano:98,Schnaiter:nanograins:99}. Laboratory measured
spectra of the solid state materials all
resemble the global appearance of the observed UIR spectrum. Looking 
in detail, however, they do not match the observed peak positions,
the observed widths and the observed profiles. In addition, these
grains would be generally too cool to emit efficiently in the mid-IR. In
summary, to date the molecular carrier of the so-called 7.8 \mum \,
component remains unidentified albeit that it likely has a highly
aromatic character.

\subsubsection{The 8.2 \mum\, feature}
\label{feat82}

Two Post-AGB stars (CRL~2688 and IRAS~13416) in our sample show a
peculiar IR spectrum. Instead 
of a 7.7 \mum\, and 8.6 \mum\, feature, they exhibit a broad 8.2 
\mum\, feature. The other UIR bands are also slightly different. The
3.3 \mum\, feature has a similar peak position as HD~44179 (type 2 in
Tokunaga et~al. 1991). But, it is broader than in HD~44179 or in
HII regions.  In addition, these objects emit a symmetric 6.3 \mum\,
feature. Unfortunately, the 11.2 \mum\, feature is too
weak to define its profile and peak position. Both
sources show a 3.4 \mum \, band. 

There are several ways to interpret these spectra. First, the observed
spectra could be a combination of emission by PAHs, giving rise to a
(slightly modified) UIR spectrum, and by an unknown carrier which
produce exclusively the 8.2 \mum\, feature.
In this interpretation, the 7.7 and 8.6 \mum\, bands would be hidden in
the strong 8.2 \mum\, feature. Second, the carriers of the
features in these sources might have similar CH modes as PAHs but their CC
modes are different. Since dust is formed in the outflows of Post-AGB
objects, the spectra of these sources might then reflect that of
freshly synthesised PAHs, dust and intermediate compounds.

This broad 8.2 \mum\, band may well be present in some other sources
as an underlying plateau (Sect. \ref{classes}). Studies of the spatial
distribution of this plateau have shown that it is carried by a
component which is independent of that of the 6.2 and 7.7 \mum\, bands
\citep{Bregman:hiivspn:89, Cohen:southerniras:89}. Energetic arguments
suggests that the carrier of this plateau contains $\sim$400 C-atoms
and, hence, the carrier may be in the form of PAH clusters
\citep{Bregman:hiivspn:89}.

In these two Post-AGB objects, the carrier of the 8.2 \mum\, band may
also be in the form of larger grains. The dust (and gas) in these two
Post-AGB objects is very close to the central star and hence dust
particles might attain high enough temperature to emit strongly around
8 \mum. The large width of the 8.2 \mum\, band lends some credence to
a grain-like carrier for this band. Various carbonaceous materials
show an emission near 8 \mum, including HAC, QCC, coal, and partially
hydrogenated C$_{60}$ \citep{Mortera:carbon:83, Sakata:QCC:84,
  Colangeli:amorphc:95, Guillois:coaltoppn:96,
  Scott:HAC:97,Schnaiter:nanograins:99, Stoldt:c60:01}. However, the
profiles in these solid state materials peak close to 8 \mum\, and are
much broader then the 8.2 \mum\, component. These materials do provide
reasonable fits to the 8 \mum\, feature as observed toward
IRAS~22272+5435 \citep{Buss:ppn:93, Guillois:coaltoppn:96,
  Kwok:irplat:01} but cannot be the carriers of the 8.2 \mum\,
feature.

\subsection{Formation/Evolution}
\label{formevol}

It was recognised some time ago that there are two main classes of
contributors to the 7.7 \mum \, band, one peaking near 7.6 \mum \,
which is associated with HII regions, and one peaking near 7.8 \mum,
associated with planetary nebulae \citep{Bregman:hiivspn:89,
  Cohen:southerniras:89}.  In this paper, we reported three main
classes. These classes are indeed related with the type of object (see
Sect. \ref{classes}). In general, HII regions, RNe, non-isolated
Herbig AeBe stars and the extragalactic sources form one class (\ca)
with dominant bands at $\sim$ 6.22 \mum, $\sim$ 7.6 \mum\, and $\sim$
8.6 \mum. The local radiation field G$_0$ in the latter sources
  ranges from 3E2 to 7E6. Isolated Herbig AeBe stars, PNe, HD44179,
HR4049 - with a local radiation field G$_0$ between 6E4 and
  2E7 -  form a second class (\cb) with dominant bands in the range of
6.24-6.28 \mum, longwards of 7.7 \mum \,and longwards of 8.62 \mum.
Two peculiar Post-AGB stars form a third class (\cc) with a
6.3 \mum\, and a 8.22 \mum \, feature; with a local radiation
  field G$_0$ of 5E3 in one of these two stars. Two Post-AGB stars
and one PN belong as well to class \ca.  As discussed in the previous
section, class \ca\, is probably built up by nitrogen substituted or
complexed PAHs while class \cb\, seems to be dominated by pure PAHs in
their dust collection. The PAH spectrum in the 6 to 9 \mum\, region
apparently reflects local physical conditions and the accumulated
effect of processing from the formation sites in the AGB or post-AGB
phases to the ISM.

The two main PAH classes identified through the peak position of
their CC modes - classes \ca\, and \cb\, - imply two distinct
histories. Because these two PAH classes are connected to classes of
astronomical objects, these histories are likely locally
determined. Hence, we interpret class \cb\, as the PAHs formed in the
stellar ejecta presumably through chemical processes similar to
terrestrial soot formation. Extensive laboratory studies and
theoretical calculations have shown that in that case highly
condensed, "pure" carbon PAHs are the favoured molecular
intermediaries in the dust condensation route
\citep{Frenklach:dustform:90, Frenklach:form:89,
  Cherchneff:formation:92}. The PAHs in class \ca, on the other hand,
likely represent a modification of those in class \cb\, in the harsh
environment of the interstellar medium
\citep{Allamandola:modelobs:99}. Interstellar PAHs could be
substantially chemically processed in the warm gas of strong shock
waves. Alternatively, energetic processing through UV and/or cosmic
rays may lead to some modification of the PAH structure
\citep{Strazulla:irrpah:95, Bernstein:pah+ice:99,Ricca:nitrogeninpahs:02,
  Ricca:nitrogenincations:02}. 

 \citet{Hony:oops:01} concluded from their study of the CH
  out-of-plane bending modes that evolution is an integral aspect of
  the life of interstellar PAHs. Specifically, the spectra of PNe
  point towards the presence of large ($\sim$ 150 C-atoms) compact PAH
  with regular molecular edge structures. In contrast, HII regions are
  dominated by highly irregular molecular edge structures. In their
  view, these molecular differences are driven by energetic processing
  in the harsh condition of the ISM of PAHs initially formed in
  stellar outflows. Our study of the CC modes complements this view.
  The spectral variation in the CC modes in these very different
  environments similar attest to the chemical evolution of the family
  of PAHs.

\section{Summary}
\label{summary}

The most striking aspect of the features in the 6--9 \mum\, region is
their variability.  All features shift in peak position from
source to source, show different profiles and each seems to be
composed of several subfeatures. Moreover, the variations in the 6.2
and 7.7 \mum\, bands seem to be 
correlated with each other.  In addition, these variations depend on the
type of source considered and apparently reflect local physical
conditions or the accumulated effect of processing from the formation
sites in the AGB or post-AGB phases to the ISM.  In particular, the
sources with a profile \ca\, 6.2 \mum\, feature have a "7.7" feature peaking
at 7.6 \mum, while for those with a component \cb\, 6.2 \mum\, feature, the
"7.7" feature peaks longwards of 7.7 $\mu$m.  The class \cc\, objects,
with a 6.2 \mum\, feature peaking at 6.3 \mum\, do not show a "7.7"
feature but instead show a broad emission band at 8.2 $\mu$m.  
 These variations stand in marked contrast to the behaviour of 3.3
and 11.2 \mum \, bands whose profiles and peak positions are quite
invariable \citep[][Peeters et~al. 2002, in prep.; van Diedenhoven
et~al. 2002, in prep.]{Hony:oops:01}.

We have summarised laboratory data and quantum chemical calculations
for the 6--9 \mum\, region. We attribute the observed 6.2 \mum\,
profile and peak position to the combined effect of a PAH family and
anharmonicity with pure PAHs representing the 6.3 \mum\, component and
substituted/complexed PAHs the 6.2 \mum\, component. The 7.6 \mum\,
component is well reproduced by both pure and substituted/complexed
PAHs but the 7.8 \mum\, component remains an enigma. In addition, the
exact identification of the 8.22 \mum\, feature remains unknown.

The observed spectral variations in the CC modes are coupled to the
astronomical characteristics of the sources (object type). We find
this strong support for the presence of a family of PAHs whose
composition and/or emission characteristics are sensitive to the local
physical conditions. An analysis of the CH out-of-plane bending modes
in a similar sample has drawn essentially the same conclusion.
Apparently, the interstellar PAH family is readily processed in space
environments.

The past decade has witnessed great experimental and theoretical
progress in understanding the spectroscopic properties of PAHs under
interstellar conditions.  However, {\bf the} new observational data
presented here pose{\bf s} significant new questions concerning the nature of
the band carriers, questions whose answers will yield insight into the
nature of the emitters and history of the emission zones.

\begin{acknowledgements}
We would like to thank the referee Dr. L. Verstraete whose comments
have helped to improve the paper.
EP and SH acknowledges the support from an NWO program subsidy (grant number
783-70-000 and 616-78-333 respectively). CVK is a Research Assistant of the
Fund for Scientific Research, Flanders. The laboratory work was supported
by NASA's Laboratory Astrophysics Program (grant number
344-02-04-02). IA$^3$ is a joint development of
the SWS consortium. Contributing institutes are SRON, MPE, KUL and the
ESA Astrophysics Division. 
This work was supported by the Dutch ISO
Data Analysis Center (DIDAC). The DIDAC is sponsored by SRON, ECAB,
ASTRON and the universities of Amsterdam, Groningen, Leiden and
Leuven. This research has made use of the SIMBAD database, operated at
CDS, Strasbourg, France and the NASA/IPAC
Extragalactic Database (NED), operated by JPL, CalTech and NASA. 
\end{acknowledgements}

\bibliographystyle{aa}
\bibliography{/arg1/users/peeters/THESIS/References/aamnem99,/arg1/users/peeters/THESIS/References/ALL,/arg1/users/peeters/THESIS/References/PAH}

\end{document}